\documentclass[fleqn,usenatbib]{mnras}

\usepackage{newtxtext,newtxmath}

\usepackage[T1]{fontenc}

\DeclareRobustCommand{\VAN}[3]{#2}
\let\VANthebibliography\thebibliography
\def\thebibliography{\DeclareRobustCommand{\VAN}[3]{##3}\VANthebibliography}

\usepackage{graphicx}	
\usepackage{amsmath}	
\usepackage{xcolor}
\usepackage{array}

\newcommand{\pcm}{~cm$^{-2}$}	
\newcommand{\pcmm}{~cm$^{-3}$}	
\newcommand{\kms}{~km\,s$^{-1}$}

\title[HCO$^{+}$, HCN, HNC, $N_2H^+$ and NH$_{3}$ deuterium fractionation in HMSF regions]{Variations of the HCO$^{+}$, HCN, HNC, N$_2$H$^+$ and NH$_{3}$ deuterium fractionation in high-mass star-forming regions}

\author[A. G. Pazukhin et al.]{
A. G. Pazukhin,$^{1,2}$\thanks{E-mail: a.pazuhin@ipfran.ru}
I. I. Zinchenko,$^{1}$\thanks{E-mail: zin@ipfran.ru}
E. A. Trofimova,$^{1}$
C. Henkel$^{3,4}$
and D. A. Semenov$^{5,6}$
\\
$^{1}$Federal Research Center A.V. Gaponov-Grekhov Institute of Applied Physics of the Russian Academy of Sciences, \\ 46 Ul’yanov str., Nizhny Novgorod 603950, Russia\\
$^{2}$National Research Lobachevsky State University of Nizhny Novgorod, 23 Gagarin Ave, Nizhny Novgorod 603950, Russia\\
$^{3}$Max-Planck-Institut f\"ur Radioastronomie, Auf dem H\"ugel 69, Bonn 53121, Germany\\
$^{4}$Astronomy Department, King Abdulaziz University, PO Box 80203, Jeddah 21589, Saudi Arabia\\
$^{5}$Max-Planck-Institut f\"ur Astronomie, K\"onigstuhl 17, Heidelberg 69117, Germany\\
$^{6}$Department of Chemistry, Ludwig Maximilian University, Butenandtstr. 5-13, Munich 81377, Germany
}

\date{Accepted XXX. Received YYY; in original form ZZZ}

\pubyear{2023}

\begin{document}
\label{firstpage}
\pagerange{\pageref{firstpage}--\pageref{lastpage}}
\maketitle

\begin{abstract}
We use spectra and maps of the $J=1-0$ and $J=2-1$ DCO$^{+}$, DCN, DNC, $\rm N_2D^+$ lines and $1_{11}-1_{01}$ ortho- and para-NH$_{2}$D lines, obtained with the IRAM-30m telescope, as well as observations of their hydrogenated isotopologues to study deuteration processes in five high-mass star-forming regions.
The temperature was estimated from CH$_3$CCH lines, also observed with the IRAM-30m telescope, and from NH$_3$ lines, observed with the 100-m radio telescope in Effelsberg, as well as using the integrated intensity ratios of the $J=1-0$ H$^{13}$CN and HN$^{13}$C lines and their main isotopologues. 
Applying a non-LTE radiative transfer model with RADEX, the gas density and the molecular column densities were estimated. 
D/H ratios are 0.001--0.05 for DCO$^{+}$, 0.001--0.02 for DCN, 0.001--0.05 for DNC and 0.02--0.4 for NH$_{2}$D. The D/H ratios decrease with increasing temperature in the range of 20--40~K and slightly vary at densities $n(\rm H_2) \sim 10^4-10^6$\pcmm. 
The deuterium fraction of $\rm N_2H^{+}$ is 0.008--0.1 at temperatures in the range of 20--25~K and at a density of $\sim 10^5$\pcmm. We also estimate relative abundances and find $ \sim 10^{-11}-10^{-9}$ for DCO$^{+}$ and DNC, $ \sim 10^{-11}-10^{-10}$ for $\rm N_2D^+$ and $ \sim 10^{-10}-10^{-8}$ for NH$_{2}$D. The relative abundances of these species decrease with increasing temperature. However, the DCN/H$_2$ ratio is almost constant ($\sim 10^{-10}$). 
The observational results agree with the predictions of chemical models (although in some cases there are significant differences).

\end{abstract}




\section{Introduction}

Our knowledge of high-mass star formation (HMSF) and early evolution of massive stars is still far from being satisfactory. The HMSF regions are rare and are located at large distances, hence understanding the involved physical and chemical processes is important~\citep[e.g.,][]{Tan14}. One of the questions is related to deuterium fractionation in these regions.

The deuterium fraction $D_{\rm frac}$ is a ratio of abundances of a deuterated molecule and its hydrogenated counterpart. 
The observed abundance of deuterated molecules in star-forming regions is higher than the initial D/H ratio $\sim 10^{-5}$~\citep{Oliveira03}. 
The abundance of deuterium in interstellar molecules increases because the forward reaction (deuteration) occurs without a thermal barrier, while the reverse reaction (removal of D) has an energy barrier~\citep[e.g.,][]{Turner}:
\begin{eqnarray}
\label{react1}
{\rm H_3^+} + {\rm HD} & \rightleftarrows & {\rm H_2D^+} \ \, \ + {\rm H_2} + 230{\rm~K}, \\
\label{react2}
{\rm CH_3^+} + {\rm HD} & \rightleftarrows & {\rm CH_2D^+} + {\rm H_2} + 370{\rm~K}, \\ 
\label{react3}
{\rm C_2H_2^+} + {\rm HD} & \rightleftarrows & {\rm C_2HD^+} + {\rm H_2} + 550{\rm~K}.
\end{eqnarray}

Reaction~(\ref{react1}) is efficient at temperatures of $\sim$10--30~K, while reactions~(\ref{react2}),~(\ref{react3}) are efficient at temperatures up to $\sim$80~K.
At densities above 10$^5$\pcmm and temperatures below 10~K, the freezing out of gaseous species onto grain surfaces, such as CO, also causes 
deuteration enhancement due to the depletion of reactions between CO and the ions in~(\ref{react1}),~(\ref{react2}) and (\ref{react3})~\citep[e.g.,][]{caselli99}.

There are chemical models that describe the formation of deuterated molecules~\citep{Turner,Roueff07,Albertsson,Sipila15,Sipila19}. $\rm N_2D^+$, DCO$^{+}$, DNC and NH$_{2}$D are mainly formed via the low temperature pathway (\ref{react1}), while DCN can be formed at high temperatures via the reactions~(\ref{react2}) and~(\ref{react3}).
The reaction~(\ref{react1}) with H$^{+}_3$ at temperatures around 30~K begins to proceed in the reverse direction, and the deuterium fraction decreases with increasing temperature.

This effect of deuterium fractionation is observed in low-mass as well as in high-mass star-forming regions. The $D_{\rm frac}$ varies with temperature and can be used as an evolutionary indicator. 
For instance, \cite{Crapsi05} have carried out a survey of $\rm N_2H^+$ and $\rm N_2D^+$ towards 31 low-mass starless cores using the IRAM-30m telescope. They recognized that high deuterium fractionation of $\rm N_2H^+$ characterise the most evolved, or "prestellar", starless cores. For massive star-forming regions, \cite{Fontani11} observed rotational transitions of $\rm N_2D^+$ and $\rm N_2H^+$ and derived the deuterium fraction in 27 cores, with the IRAM-30m telescope. They concluded that the $\rm N_2D^+$-to-$\rm N_2H^+$ column density ratio can be used as an evolutionary indicator. Moreover, for the same regions in \cite{Fontani14, Fontani15} they have estimated NH$_2$D/NH$_3$ and DNC/HNC. $D_{\rm frac}$(NH$_3$) is on average above 0.1 and does not change significantly with evolutionary phase. For DNC/HNC, they have found no statistically significant differences among the three evolutionary groups of objects such as high-mass starless cores (HMSCs), high-mass protostellar objects (HMPOs) and  ultracompact HII regions (UCHIIs). Additionally, \cite{Sakai12} found that the DNC/HNC ratio does not depend only on the current kinetic temperature towards 18 massive clumps (IRDCs and HMPOs), by using the Nobeyama Radio Observatory 45 m telescope. With a chemical model, they suggested that the DNC/HNC ratio also depends on history in their starless-core phase, such as its duration time. 

\cite{Gerner15} observed a sample of 59 high-mass star-forming regions with different evolutionary phases: starting with IRDCs via HMPOs to hot molecular cores (HMCs) and finally UCHIIs regions. They found that the D/H ratios of DNC, DCO$^+$, and $\rm N_2D^+$ show decreasing trends with evolutionary stages, despite high standard deviations of ratios within individual stages. However, DCN/HCN shows maxima in the HMC phase. Also $\rm N_2D^+$ was only detected in a few IRDCs and HMPOs. \cite{Trofimova20} have undertaken a survey of 60 massive star forming regions in DCN, DNC, DCO$^+$, $\rm N_2D^+$, by using the 20-m Onsala radio telescope. The $\rm N_2D^+$ was detected only in two sources, other deuterated molecules were detected in about 1/3 of the sources. They have found that the abundances relative to H$_2$ of DCN and DNC and the DCN/HCN ratio are almost constant at temperatures of $\sim$15--55~K, while DCO$^+$/H$_2$ decreases with increasing temperature. 
Using the Mopra-22m and the IRAM-30m telescopes, \cite{Wienen2021} observed NH$_2$D at 86 and 110 GHz towards over 900 high-mass clumps discovered by the APEX Telescope Large Area Survey of the Galaxy (ATLASGAL). They did not find a correlation between the NH$_3$ deuteration and evolutionary relevant physical tracers such as rotational temperature.

There are few mapping surveys to study the deuterium fractionation. For instance, using the IRAM-30m telescope, \cite{Feng19} have imaged two high-mass protostellar clumps that show different evolutionary stages in IRDC G28.34+0.06. They have found that the deuteration of $\rm N_2H^+$ is more efficient than that of HCO$^+$, HCN, and HNC. The deuterations are favoured towards the chemically younger clump with its colder and denser environment. The NH$_2$D abundance is almost independent of environmental differences. 
\cite{Pillai12} obtained maps of the ortho-$\rm H_2D^+$(1$_{10}$–1$_{11}$) and $\rm N_2H^+$(4–3) lines with the James Clerk Maxwell Telescope (JCMT), and $\rm N_2D^+$(3--2) and dust continuum  with the Submillimeter Array (SMA) in the DR21 filament of Cygnus X. The H$_2$D$^+$ emission is widely associated with dust emission and $\rm N_2D^+$, however the H$_2$D$^+$ peaks are offset from the dust and $\rm N_2D^+$(3--2) peaks. \cite{Tan13} obtained maps of $\rm N_2D^+$(3--2) and DCO$^+$(3--2) emissions from four IRDCs using ALMA. In addition, \cite{Coutens14}  detected the HDO and H$_{2}^{18}$O transitions with the Herschel/HIFI instrument, the IRAM 30-m telescope, and the CSO towards the HMSF region G34.26+0.15. The radial variation of the deuterium fraction was determined using a 1D non-LTE radiative transfer code. The HDO/H$_2$O ratio is estimated to be $\sim 3.5-7.5\times10^{-4}$ in the hot core ($\sim$200\,K) and $\sim 1.0-2.2\times10^{-4}$ in the colder envelope ($\sim$100\,K). In a recent study, \cite{Redaelli21} performed ALMA mapping observations of the continuum emission at 0.8 mm and of the ortho-$\rm H_2D^+$(1$_{10}$–1$_{11}$) towards the two infrared-dark massive clumps. They found that ortho-$\rm H_2D^+$ is an ideal tracer of cold ($\sim10$~K) and dense ($\sim10^6$\pcmm) gas.
 
 In this work, we study the physical and chemical conditions of high-mass star-forming regions, using observations with the IRAM-30m radio telescope and the 100-m radio telescope in Effelsberg. We investigate the spatial distribution of abundances, temperature and density. We also wish to compare observational results with model predictions.
We use the integrated intensity ratios of the $J=1-0$ HCN and HNC lines and their $\rm ^{13}C$ bearing isotopologues as temperature indicator.
Assuming optically thin molecular emission we estimate the gas density and the molecular column densities using a non-LTE radiative transfer model applying the RADEX code. 
We derive and discuss abundances of the deuterated molecules DCO$^{+}$, DCN, DNC, $\rm N_2D^+$ and NH$_{2}$D as functions of gas properties such as temperature and density. We also discuss the spatial distribution of molecules. To obtain abundances, we derive H$_2$ column densities from $850\, \micron$ SCUBA dust emission. Previous works were mainly based on single-dish pointing surveys, while our study is one of the first including maps to study the deuterium fractionation in HMSF regions. 

The paper is organized as follows. Section \ref{obs} describes the observations and data reduction. The results and discussions are presented in Section \ref{res} and \ref{dis}. A summary of the main conclusions is presented in Section \ref{con}.

\section{Observations and data reduction}\label{obs}

\subsection{Observations at the 30-m radio telescope of the Institut de Radioastronomie Millim{\'e}trique (IRAM)}

In September 2019, with the 30-m radio telescope of the Institut de Radioastronomie Millim{\'e}trique (IRAM), we observed five massive star forming regions at wavelengths of 2 and 3--4~mm (in the framework of the project 041-19). The sources are selected from the sample of a previous survey conducted at Onsala \citep{Trofimova20}, possessing a relatively strong emission in the lines of deuterated molecules and having different gas temperatures. The list of sources is given in Table~\ref{tab1}.  The source position list is mainly based on a galactic H$_2$O maser catalogue~\citep{Palagi93,Valdettaro01,Ladeyschikov19}. The L1287 source position is associated with the IRAS~00338+6312. For S187 the central position corresponds to the submillimetre dust emission peak and the N$_2$H$^+$ peak \citep{Zinchenko09} associated with the massive pre-main-sequence star S187H$\alpha$ \citep{Zavagno94}. Table~\ref{tab2} contains the list of the observed molecular lines with some spectroscopic parameters.
Only one source, DR21(OH), was observed at $\sim$110~GHz in the para-NH$_{2}$D line due to limited observing time. Transition frequencies and upper level energies are taken from The Cologne Database for Molecular Spectroscopy (CDMS)\footnote{\url{http://cdms.de}}~\citep{cdms}.

\begin{table*}
\centering
\caption{List of sources.} 
    \begin{tabular}{lccccc}
            \hline
            Source & RA(J2000) & Dec(J2000) & $V_{lsr}$ & $d$& Note \\
                   & $(^{h}\ ^{m}\ ^{s})$ & $(\degr\ \arcmin\ \arcsec)$ & (km s$^{-1}$) & (kpc) & \\
            \hline
            L1287     & 00:36:47.5 & 63:29:02.1 & -17.7 & 0.9$^a$ & G121.30+0.66, IRAS~00338+6312\\
            S187     & 01:23:15.4 & 61:49:43.1 & -14.0 & 1.4$^b$  & G126.68–0.81, IRAS~01202+6133 \\
            S231     & 05:39:12.9 & 35:45:54.0 & -16.6 & 1.6$^c$  & G173.48+2.45, IRAS~05358+3543 \\
            DR21(OH) & 20:39:00.6 & 42:22:48.9 &  -3.8 & 1.5$^d$  & G81.72+0.57 \\
            NGC7538  & 23:13:44.7 & 61:28:09.7 & -57.6 & 2.7$^e$  & G111.54+0.78, IRAS~23116+6111 \\
            \hline
        
        \multicolumn{6}{l}{Distances to sources are quoted from $^a$\cite{Rygl10}, $^b$\cite{Russeil07}, $^c$\cite{Burns15}, $^d$\cite{Rygl12}, }\\
        \multicolumn{5}{l}{$^e$\cite{Moscadelli09}}
    \end{tabular}    
    \label{tab1}
\end{table*}

\begin{table}
\centering
\caption{Observed molecular lines.} 
\begin{tabular}{lcrr}
        \hline
        Molecule & Transition & Rest frequency & $E_{u}/k$ \\
         &  & (MHz) & (K) \\
        \hline
        NH$_3$ & $ (J,K)=(1,1)$ & 23694.495 & 23.4 \\
               & $ (J,K)=(2,2)$ & 23722.634 & 64.9 \\
               & $ (J,K)=(3,3)$ & 23870.128 & 123.5 \\
        DCO$^{+}$ & $ J=1-0$ & 72039.354 & 3.5  \\
		       & $ J=2-1$ & 144077.319 & 10.4  \\ 
	DCN & $ J=1-0, F=2-1$ & 72414.694 & 3.5  \\ 
		& $ J=2-1, F=2-1$ & 144828.002 & 10.4  \\ 
	DNC & $ J=1-0$  & 76305.699 & 3.7  \\ 
	      & $ J=2-1$  & 152609.744 & 10.9 \\
        $\rm N_{2}D^{+}$ & $ J=1-0$  & 77109.243 & 3.7  \\ 
		          & $ J=2-1$  & 154217.011 & 11.1 \\
        CH$_{3}$CCH & $ J_K=5_{3}-4_{3}$ & 85442.601 & 77.3 \\
                    & $ J_K=5_{2}-4_{2}$ & 85450.766 & 41.2 \\
                    & $ J_K=5_{1}-4_{1}$ & 85455.667 & 19.5 \\
                    & $ J_K=5_{0}-4_{0}$ & 85457.300 & 12.3 \\
        NH$_{2}$D  
                    ortho &  $ J_{K_a K_c}=1_{11}-1_{01},F=2-2$ & 85926.278 & 20.7 \\
        \hspace{3em} para &  $ J_{K_a K_c}=1_{11}-1_{01},F=2-2$ & 110153.594 & 21.3 \\        
        H$^{13}$CN & $ J=1-0, F=2-1$ & 86339.921 & 4.1 \\
        H$^{13}$CO$^{+}$ & $ J=1-0$ & 86754.288  & 4.2 \\
        HN$^{13}$C & $ J=1-0$ & 87090.825 & 4.2 \\
        HCN & $ J=1-0, F=2-1$ & 88631.602 & 4.3 \\
        HCO$^{+}$ & $ J=1-0$ & 89188.525 & 4.3 \\
        HNC & $ J=1-0$ & 90663.568 & 4.4 \\
        CH$_{3}$CCH & $ J_K=9_{3}-8_{3}$ & 153790.772 & 101.9 \\
                    & $ J_K=9_{2}-8_{2}$ & 153805.461 & 65.8 \\
                    & $ J_K=9_{1}-8_{1}$ & 153814.276 & 44.1 \\
                    & $ J_K=9_{0}-8_{0}$ & 153817.215 & 36.9 \\
        \hline
    \end{tabular}
    \label{tab2}
\end{table}

The full beam width at half maximum at the observed frequencies ranged from $\sim36\arcsec$ to $\sim17\arcsec$.  Antenna temperatures $T^*_{\rm A}$ were converted to the main beam brightness temperature $T_{\rm mb}$, using the main beam efficiency $B_{\rm eff}$, which was determined by the Ruze's formula in accordance with the IRAM recommendations\footnote{\url{https://publicwiki.iram.es/Iram30mEfficiencies}} and ranged from 0.72 to 0.82. The minimum system noise temperatures were $\sim 100$~K in the 3~mm range and $\sim 200$~K in the 2~mm range.

Observations were carried out in the on-the-fly (OTF) mode over a mapping area of
a $200\arcsec\times200\arcsec$ in total power mode. The reference position was chosen with a shift of $10\arcmin$ in right ascension. In some extended sources, {i.e. DR21(OH) and NGC7538,} two partially overlapping areas were observed. The pointing accuracy was checked periodically by observations of nearby continuum sources.

\subsection{Observations at the Max-Planck-Institut f{\"u}r Radiostronomie {with the} Effelsberg 100-m radio telescope}
{On 9 December 2019 we observed with the 100-m telescope near Effelsberg (Germany) the H$_{2}$O maser} transition at a frequency of 22~GHz, as well as the ammonia inversion lines {$(J,K)=\,$}(1,1), (2,2) and (3,3). The full beam width at half maximum was $\sim40^{\prime\prime}$.
The measurements were carried out by the method of continuous mapping using a $K$-band receiver in a secondary focus with a dual bandwidth of 300~MHz, including the H$_{2}$O lines in one band and NH$_3$ in the other band. $5\arcmin \times 5\arcmin$ maps were obtained at a scanning rate of 20$\arcsec$ per second in right ascension; intervals between scans were 15$\arcmin$. The reference position was shifted by +15$\arcmin$ in azimuth. Weather conditions included light rain with low wind speeds ($\sim$2~m\,s$^{-1}$).

The results are presented in the main beam temperature scale $T_{\rm mb}$.
The source NGC7027 was used for calibration with a flux density $S_{\rm Ott}$ of 4.7~Jy at 22~GHz, taking into account the annual change since 1990~\citep{Ott94}.

\subsection{Archival data}
Ammonia emission from L1287 was taken from observations with the Effelsberg-100m telescope in 1995 \citep{Zinchenko97}. 
We used ammonia observations towards the interstellar filament WB~673 with the Effelsberg-100m telescope in 2019 from \cite{Ryabukhina}. 
Also we used data from the KEYSTONE survey with the 100~m Green Bank Telescope mapping ammonia emission across giant molecular clouds (Cygnus~X~North, NGC7538) from \cite{Keown19}. 
The $\rm N_2H^+$ column densities were adopted from observations with the 20-m OSO telescope and the 15-m SEST telescope~\citep{Pirogov03}. 
The continuum data are obtained with the James Clerk Maxwell telescope (JCMT)-SCUBA at 850~$\micron$ \citep{scuba} as the dust distribution indicator.

\subsection{Data reduction}\label{datared}
The GILDAS/CLASS software\footnote{\url{http://www.iram.fr/IRAMFR/GILDAS}} was used for data reduction. 
All datasets were smoothed to the same spatial resolution of 40$\arcsec$. The spectra were fitted with Gaussian profiles using the \texttt{LMFIT} package \citep{lmfit}. In the analysis, integrated intensity was obtained from the Gaussian profile area with their errors as the fit errors.  
For the spectra with hyperfine structure we assume that the widths of all components are equal, and the spacings between them are known. This allows us to determine the optical depth of the main group of hyperfine components, $\tau$. Assuming that the ratios of hyperfine components correspond to LTE conditions, 
the optical depth of the main line can be determined from the ratio of the observed main and satellite line intensities:
 \begin{equation}
    \frac{T_{\rm mb}({m})}{T_{\rm mb}({s}) }=\frac{1-\exp(-\tau(m))} {1- \exp(-a\tau(m))},
\label{eq:tau}
\end{equation}
where $T_{\rm mb}$ is the main-beam temperature, $a$ is the ratio of the satellite and main line strengths using the statistical weights. We adopted $a = 0.28$ and $0.22$ for the inner and outer satellites of NH$_3(1, 1)$ and o-NH$_2$D(1$_{11}$--1$_{01}$) respectively. For H$^{13}$CN and DCN, we used $a = 0.6$ for the $F=1-1$ and $F=2-1$ hyperfine components, and $a = 0.2$ for the $F=0-1$ and $F=2-1$ components.

It should be noted that two velocity components at $\sim -4$ and $\sim 0$~\kms are observed in the source DR21(OH)~\citep[see details in][]{Schneider}. In the reduction, the components were separated, and only the $\sim -4$~\kms component has been used for the analysis, since it is stronger and is detected throughout the source.

 Packages \texttt{NumPy} \citep{numpy}, \texttt{LMFIT} \citep{lmfit}, \texttt{astropy} \citep{astropy}, \texttt{matplotlib} \citep{matplotlib}, and \texttt{SciPy} \citep{scipy} were used for the fitting, numerical calculations and plotting. 

\section{Results}\label{res}

\subsection{Maps and spectra}
In Figs.~\ref{fig:maps_int_121_30} and~\ref{fig:maps_int} of the Appendix we show the integrated intensity maps to compare the spatial distribution of molecules. The dust and thus the gas column density distribution are represented by the 850\,$\micron$ SCUBA continuum emission. 
In general, all hydrogenated molecules show emission peaks at a position coincident with the main dust emission peak and the IRAS source position. The deuterated molecules present various distributions. DCN, unlike DNC, DCO$^+$ and NH$_2$D, shows emission peaks consistent with the hydrogenated isotopologues. In DR21(OH) and NGC7538, the DCN emission is stronger than DCO$^+$ and DNC, but in L1287, S231 and S187, DCO$^+$ provides the strongest emission from a deuterated molecule. Notably, in S187, the NH$_2$D line demonstrates stronger emission by a factor of $\sim 3$ than NH$_3$, but it is located at the edge of the map.
Additionally, the $\rm N_2D^+$ emissions are weak and not detected in NGC7538. 

\begin{figure*}
    \centering \includegraphics[width=\linewidth]{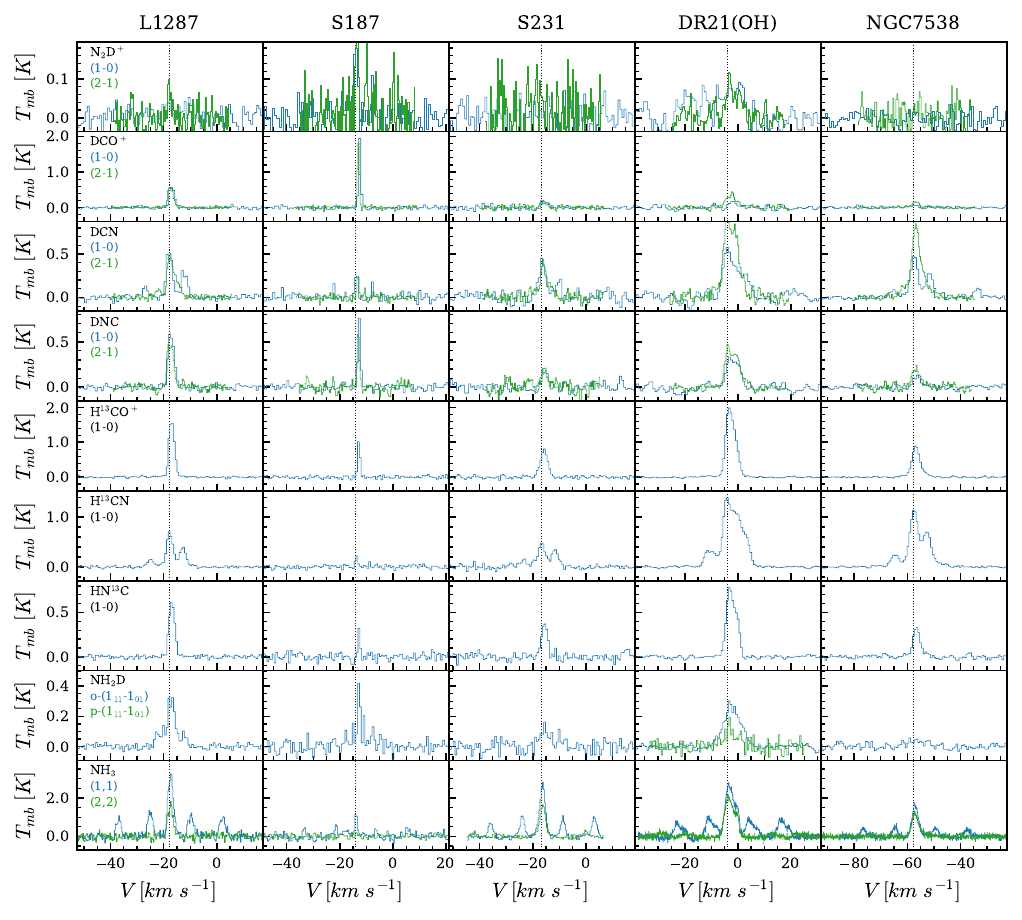}\\
    \caption{Spectra extracted at the main dust emission peak from 850\,$\micron$ SCUBA data. Source names are shown at the top of each column. Transitions are shown in the upper left corner of the first column. The system velocity is shown as a gray dashed line in each panel.}
    \label{fig:spectra}
\end{figure*}

In Fig.~\ref{fig:spectra} we show the spectra extracted at the 850\,$\micron$ main dust continuum peak. For HCO$^{+}$, HCN and HNC, the  $\rm ^{12}C$ lines are affected by self-absorption, thus $\rm ^{13}C$ lines are analysed in this paper. 
In DR21(OH) both velocity components mentioned above (see sect.~\ref{datared}) can be seen. The line widths are from $\sim$1\kms, towards S187, to $\sim$3\kms, towards DR21(OH).  In S187 the line width is comparable to the velocity resolution. In general, the hydrogenated molecular lines show stronger emission by a factor of $\sim 2$, but in S187, to the contrary, the deuterated isotopologue lines are stronger.
In addition, we also estimated the main line optical depth from the hyperfine structure of DCN(1--0) and H$^{13}$CN(1--0). The optical depths at the emission peaks were found to be low, $\tau<<1$. Furthermore, the main line optical depths of o-NH$_2$D(1$_{11}$--1$_{01}$) and NH$_{3}$(1,1) at the emission peaks are $\tau\sim1$. 

As a future perspective, we plan to investigate the kinematics and dynamics of the gas.

\begin{figure*}
    \centering \includegraphics[width=\linewidth]{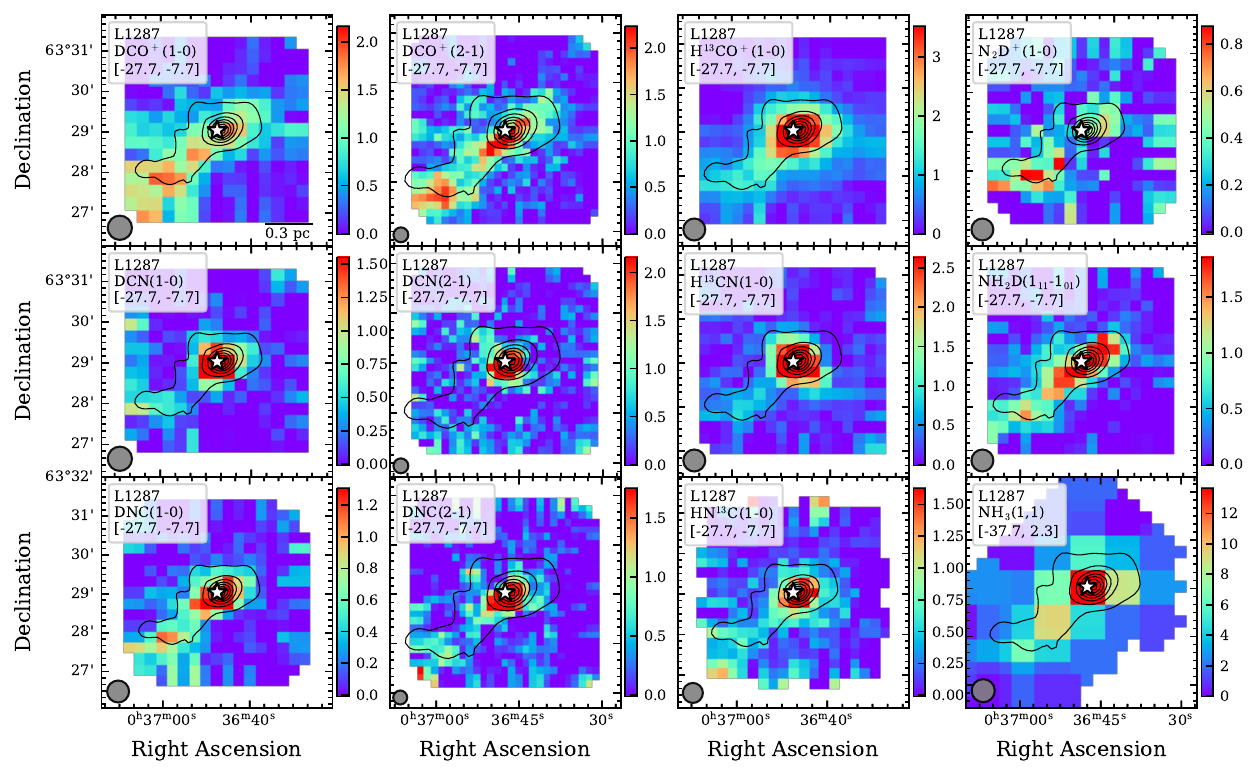}   
    \caption{The integrated intensity maps for L1287. 
    The integrated intensity is obtained by integrating the line in main beam temperature intensity units.
    The contours show the continuum emission from 850\, $\micron$ SCUBA data. The levels start from 5\% to 95\% of the peak intensity of 6.5 $\rm mJy\, beam^{-1}$ in steps of 15\%. The star-shaped marker indicates the IRAS source position. Sources, transitions  and the velocity range are shown in the upper left corner of each panel. The beam sizes are shown in the bottom left corner of each panel. A scale bar representing a linear scale of 0.3 pc is shown on the bottom-right corner of the first frame. The maps of the other sources are presented in the Appendix (Fig.~\ref{fig:maps_int}).}
    \label{fig:maps_int_121_30}
\end{figure*}

\subsection{Kinetic temperature from observations of CH$_{3}$CCH}

In~\cite{Askne84} and \cite{Bergin} it was shown that the rotational temperature of CH$_{3}$CCH gives a good estimate of the gas kinetic temperature at gas density $n\ga 10^{3-4}$~cm$^{-3}$ (transitions $J=5-4$ and $J=6-5$ were considered). This is explained by the fact that, due to the low dipole moment ($\mu$ = 0.78~D), the CH$_{3}$CCH molecule is easily thermalized under such conditions. Gas densities in our sources are above this threshold. Thus, the CH$_{3}$CCH lines in our data can be a good gas kinetic temperature indicator. Rotational (and, accordingly, kinetic) temperature is determined from the rotation diagram method. It is assumed here that the emission is optically thin and the background radiation can be neglected.

The rotational diagrams were constructed using the $J=5-4$ and $J=9-8$ transitions of the CH$_{3}$CCH molecule. To estimate the rotational temperature, we smooth the $J=5-4$ and $J=9-8$ line maps to the same angular resolution of 40$\arcsec$. The K-ladder spectra were fitted with Gaussian profiles, assuming that the widths of all components are equal, and the spacings between them are known.

\subsection{Kinetic temperature from NH$_3$ observations }

Ammonia inversion transitions have a complex hyperfine structure with several components grouped in five lines, a central one as well as inner and outer satellites (see the lowest row of spectra in Fig.~\ref{fig:spectra}).
Optical depths and rotational temperatures were determined using the methods described in~\cite{Ho83}.
The spectra were fitted with Gaussian profiles. In the {$(J,K)=\,$}(1,1) transition line widths of all hyperfine components were assumed to be equal, and the spacings between them are known.
As described in section~\ref{datared}, the optical depth $\tau({1,1,m})$ is derived from the ratio of the observed main and satellite line intensities according to Eq.~(\ref{eq:tau}).

Thus, the rotational temperature can be obtained from the ratio of the main groups component intensities of the (1,1) and (2,2) transitions using the equation \citep{Ho83}:
\begin{eqnarray}
    T_{\rm rot} = -41.5 \Bigg/ \ln \Big[ \frac{-0.282}{\tau(1,1,m)} \\ \nonumber
    \times \ln \Big( 1-\frac{T_{\rm mb}(2,2,m)}{T_{\rm mb}(1,1,m)} \{1-\exp(-\tau(1,1,m))\} \Big) \Big].
\label{eq:trot}
\end{eqnarray}

The kinetic temperature values were obtained using the equation from~\cite{Tafalla04}:
\begin{equation}
    T_{\rm kin} = \frac{ T_{\rm rot}}{ 1- \frac{T_{\rm rot}}{41.5} \ln \left[ 1+ 1.1 \exp \left( -\frac{16}{T_{\rm rot}} \right) \right] }.
\label{eq:tkin}
\end{equation}

\subsection{Kinetic temperature from the integrated intensity HCN/HNC ratio}\label{hcn_hnc}

\begin{figure*}
\begin{minipage}{0.45\linewidth}
        \centering \includegraphics[width=\linewidth]{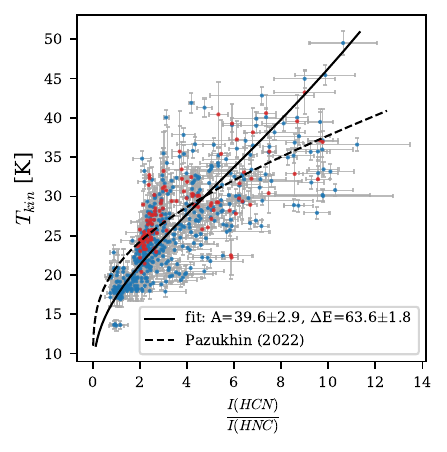}
    \end{minipage}\hfill
    \begin{minipage}{0.45\linewidth}
        \centering \includegraphics[width=\linewidth]{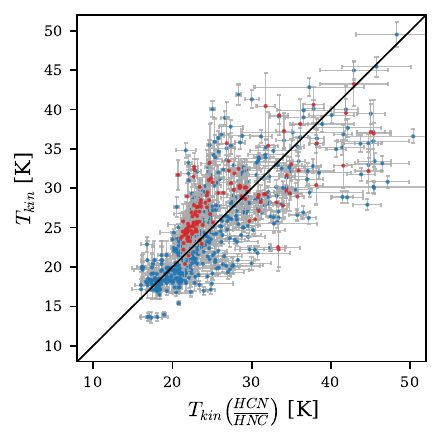}
    \end{minipage}
    \caption{ The kinetic temperature derived from NH$_3$~(blue) and CH$_3$CCH~(red) versus the integrated intensity ratio HCN/HNC (left panel) and the T$_{\rm kin}$(HCN/HNC) (right panel). The fitting results are represented by the black curve. The black dashed curve shows the fitting results from \citet{Pazukhin}.}
    \label{fig:Tkin_compare}
\end{figure*}

It is known that the HCN/HNC abundance ratio strongly depends on the kinetic temperature ~\citep[e.g.,][]{Hirota}.
In~\cite{Hacar20} it was proposed to use the intensity ratio of the  $J=1-0$ HCN to HNC line as a temperature indicator based on observations of the integral shaped filament in Orion. Following \cite{Pazukhin}, we found a correlation between the integrated intensity ratio H$^{13}$CN/HN$^{13}$C and the kinetic temperature expressed in terms of the Boltzmann distribution (see Fig.~\ref{fig:Tkin_compare}, left panel). We use the following equation as a temperature indicator:
\begin{equation}
    \frac{\rm H^{13}CN}{\rm HN^{13}C} = 40\times \exp \left( {\frac{-64}{T_{\rm kin}}} \right). \label{eq:rat_dE} \\
\end{equation}

As can be seen in Fig.~\ref{fig:Tkin_compare}, this fit is somewhat different from that found in \cite{Pazukhin}. The difference can be explained by a larger dataset as compared to the previous work by \cite{Pazukhin}. Figure~\ref{fig:Tkin_compare}(right) shows the kinetic temperature obtained from the NH$_3$ and CH$_3$CCH transitions in comparison with the temperatures estimated from the integrated intensity ratios $J=1-0$ H$^{13}$CN and HN$^{13}$C lines and their main isotopologues. In general, the T$_{\rm kin}$(HCN/HNC) values show a good agreement with the estimates derived from the CH$_3$CCH and NH$_3$ lines in the range of 20 to 40~K with deviations of $\la 5$~K. 

We further expanded the temperature maps by combining the observational data from isotopologues H$^{13}$CN and HN$^{13}$C with observations from the main isotopologues, as suggested by~\cite{Beuther22}. In those source regions where the H$^{13}$CN or HN$^{13}$C lines become too weak, the intensity ratio of the main isotopologues is used.

Figs.~\ref{fig:maps_Tk_121_30} and~\ref{fig:maps_Tk} in the Appendix represent the temperature maps.
The HCN/HNC maps demonstrate a good agreement with the estimates derived from the CH$_3$CCH and NH$_3$ lines. The temperature gradient is clearly visible, with both low-temperature and high-temperature regions being traceable. Temperature peaks coincide spatially with both the dust continuum emission and the IR source position.  In addition, the maps extend beyond the temperature maps derived from the CH$_3$CCH and NH$_3$ lines. The T$_{\rm kin}$(HCN/HNC) of individual objects are discussed in Sect.~\ref{indobj}.

\subsection{Non-LTE analysis}

Assuming non-LTE and optically thin molecular emission, the gas density and the molecular column density can be estimated using the off-line RADEX code~\citep{Radex}. 
Energy levels, statistical weights, Einstein A-coefficients and collision rate coefficients were taken from the Leiden Atomic and Molecular Database (LAMDA)\footnote{\url{https://home.strw.leidenuniv.nl/~moldata/}}~\citep{lamda}. Also collisional data were adopted from the BASECOL database\footnote{\url{https://basecol.vamdc.eu}}~\citep{Basecol}.
For D and H isotopologues, we used collision coefficients calculated for DCO$^{+}$--H$_{2}$ and HCO$^{+}$--p-H$_{2}$ by \cite{Denis20}, HCN--p-H$_{2}$ and HNC--p-H$_{2}$ by \cite{Hernandez17}, NH$_3$--p-H$_{2}$ by \cite{Bouhafs2017}, NH$_{2}$D--p-H$_{2}$ by \cite{Daniel14} and $\rm N_2H^+$--p-H$_2$ by \cite{Balanca20}.

We determined integrated intensity ratios of the $1-0$ and $2-1$ lines of DCO$^{+}$, DCN, DNC and compared them with calculated model values from RADEX. 
We built model grids with kinetic temperatures in the range $T_{\rm kin}=5-80$~K, H$_{2}$ volume densities in the range $n(\rm H_2)=10^{3}-10^{8}$\pcmm, and total column densities of 10$^{12}$\pcm. At this column density the optical depths are low in all lines, so that line intensity ratios do not vary, if column densities were actually smaller. Note that estimates of the H$_{2}$ volume densities decrease with increasing optical depth of the $1-0$ and $2-1$ transitions. Thus, we use model intensity ratios of optically thin lines, which weakly depend on the column density at $\la 10^{13}$\pcm.  To derive $n($H$_{2})$ we estimated the following chi-squared minimum:
\begin{equation}
    \chi^2 = \left(\frac{R^{\rm obs}_{i}-R^{\rm mod}_{i}}{\sigma^R_i}\right)^2 \times \left(\frac{T^{\rm obs}_{i}-T^{\rm mod}_{i}}{\sigma^{\rm T}_i} \right)^2,
    \label{eq:label} 
\end{equation}
where $R^{\rm obs}_i$ and $R^{\rm mod}_i$ are the 2--1/1--0 ratios from observations and models, $(\sigma^{\rm R}_{i})^2$ is the sum of squared rms noise of $1-0$ and $2-1$ lines, $T^{\rm obs}_{i}$ and $T^{\rm mod}_{i}$ are the kinetic temperatures from observations and models and $\sigma^{\rm T}_i$ is the uncertainty of $T^{\rm obs}_{i}$. The volume densities $n($H$_{2})$ were found from the average value for the condition $\chi^2 + 3.84$ (95~per~cent confidence level), and the errors were found from the standard deviation.

\begin{figure*}
\begin{minipage}{0.33\linewidth}
        \centering \includegraphics[width=\linewidth]{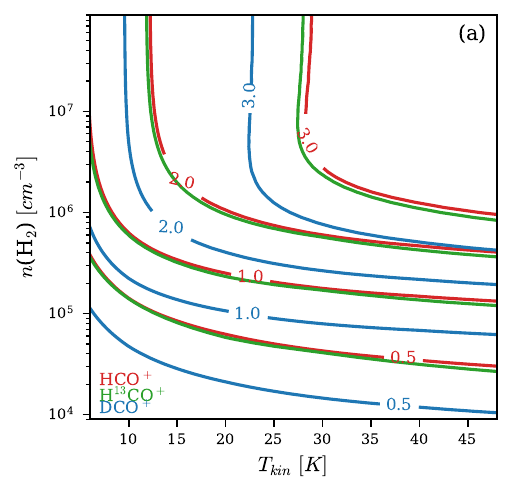}
    \end{minipage}\hfill
    \begin{minipage}{0.33\linewidth}
        \centering \includegraphics[width=\linewidth]{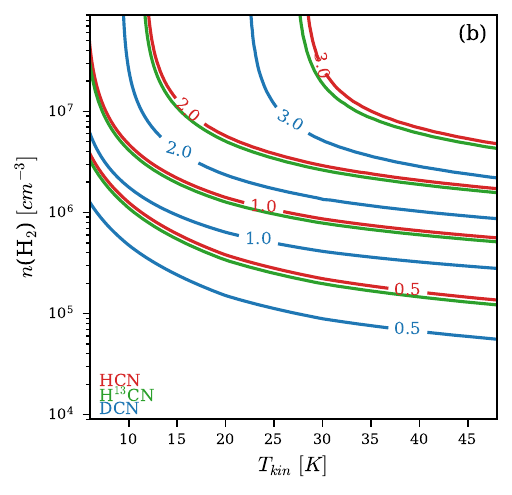}
    \end{minipage}\hfill
    \begin{minipage}{0.33\linewidth}
        \centering \includegraphics[width=\linewidth]{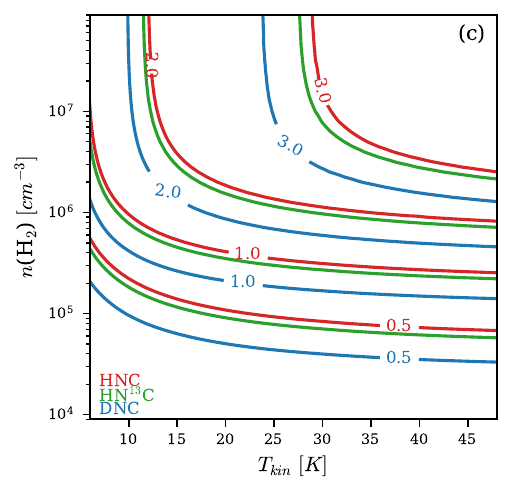}
    \end{minipage}
    \caption{(a) Curves of the constant HCO$^+$ (red lines), H$^{13}$CO$^+$ (green lines) and DCO$^+$ (blue lines) $J$=(2-1)/(1-0) intensity ratios on the $T_{\rm kin}$ -- $n\rm (H_2)$ plane from the RADEX model calculations for optically thin conditions. (b) Same for HCN, H$^{13}$CN and DCN; (c) HNC, HN$^{13}$C and DNC.}
    \label{fig:nh2_radex}
\end{figure*}

Only for DCO$^+$ and NH$_{2}$D native data files are available in the LAMDA database. In cases of DCN, DNC and N$_2$D$^+$ the data files of their hydrogenated isotopologues were used. This can lead to biases in the density estimates. Indeed, in Fig.~\ref{fig:nh2_radex} we show that the density estimates with the HCO$^+$ data file are higher than the estimates with the DCO$^+$ data file by a factor of 3. Hence, to improve the results for DCN and DNC we modified the molecular data files by substituting the frequencies, energy levels and Einstein A-coefficients for the deuterated isotopologues. The data for the $^{13}$C isotopologues were also appended. We suppose that the difference in the $n\rm(H_2)$ estimates is due not only to collision rate coefficients, but also to these parameters. The results of our substitution are shown in Fig.~\ref{fig:nh2_radex}.

As mentioned above all maps were smoothed to 40$\arcsec$ and to the same grid size. After that, using the integrated $1-0$ line intensity, the kinetic temperature and calculated $n$(H$_{2}$), the column density was obtained for each pixel in the map. In the analysis we use only the kinetic temperature derived from the integrated intensity ratios of HCN and HNC isotopologue lines (see sect.~\ref{hcn_hnc}). Since the line ratios of the $^{12}$C to $^{13}$C bearing isotopologues indicate that HCO$^{+}$, HCN, HNC lines are optically thick, the lines of their optically thin isotopologues $\rm H^{13}CO^{+},$ H$^{13}$CN, HN$^{13}$C were used. To do this, the obtained values of the column densities of optically thin isotopologues were multiplied by a coefficient obtained from the carbon isotope ratio $\frac{\rm ^{12}C}{\rm ^{13}C}=4.8\times{\rm R_{GC}}+20.8$~\citep{Yan2023}, where $\rm R_{GC}$ is the Galactocentric distance in kpc.
For $N$(NH$_3$) and $N$(NH$_{2}$D), we assumed ortho-to-para ratios of 1 and 3, respectively, which correspond to the nuclear spin statistical weights. To derive $D_{\rm frac}$(NH$_3$) and $N(\rm N_2D^+)$ values we use the mean value of $n($H$_{2})$ of DCO$^{+}$, DCN, DNC.

\subsection{H$_2$ column density}
We derived H$_2$ column densities from dust maps obtained with the SCUBA Legacy Catalogue at $850\, \micron$ and a resolution of $22.9\arcsec$. 
The maps were smoothed to $40\arcsec$ resolution to provide an optimal match with the IRAM-30m and Effelsberg-100m data.
Following \cite{Hildebrand83}, the H$_2$ column density is related to the dust emission by:
\begin{equation}
N_{\rm H_2}=\eta \frac{S_{\rm \nu}}{B_{\rm \nu}(T_{\rm dust}) \Omega \kappa_{\rm \nu} \mu m_{\rm H}},
\end{equation}
where $\eta = 100$ is the gas-to-dust ratio,
$B_{\rm \nu}(T)$ is the Planck function,
$S_{\rm \nu}$ is the observed flux, 
$\Omega$ is the beam solid angle,
$T_{\rm dust}$ is the dust temperature,
$m_{\rm H}$ is the mass of a hydrogen atom and
$\mu = 2.8$ is the mean molecular weight of the interstellar medium.
The dust opacities used were $\kappa_{\rm \nu}=1.82\, \rm cm^2\, g^{-1}$ at $850\, \micron$~\citep{Ossenkopf94}.
Then, the column density values, $N_{\rm H_2}$ in\pcm, are derived using the equation in useful units~\citep{Kauffmann08}:
\begin{equation}
N_{\rm H_2} =   2.02 \times 10^{24}
  \left( {\rm e}^{14.39/(\lambda \, T_{dust})} - 1 \right) 
   \frac{\lambda^{3}\, S_{\rm \nu}^{\rm beam}}  {\kappa_{\rm \nu} \, \theta^{2}},
\end{equation}
where $S_{\rm \nu}^{\rm beam}$ is the observed flux in $\rm mJy\, beam^{-1}$, $\theta$ is the beam size in arcsec and
$\lambda$ is the wavelength in mm.

To derive the H$_2$ column density we assumed that $T_{\rm dust}$ is 20~K based on results from \cite{Pazukhin}. They compared the gas kinetic temperature with the dust temperature, which was taken from the open database\footnote{\url{http://www.astro.cardiff.ac.uk/research/ViaLactea}} according to data from the Herschel telescope~\citep{Marsh_1,Marsh_2}. The $T_{\rm dust}$ values were in the range $\sim$18--25~K and no significant correlation between the gas and dust temperatures was found.

In addition, we estimated the mean density based on the measured N(H$_2$), assuming as the source size the FWHM of the $850\,\micron$ continuum emission and Gaussian source profile. These mean densities are given in Sect.~\ref{indobj}.

\subsection{Individual objects}\label{indobj}
\subsubsection{L1287}
\begin{figure*}
    \centering \includegraphics[width=\linewidth]{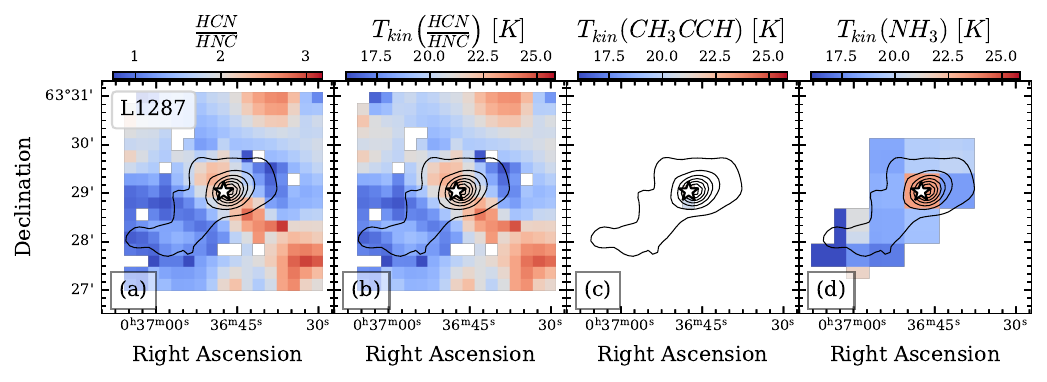}
    \caption{Maps for L1287.
    (a)~Combined integrated intensity ratios HCN/HNC and H$^{13}$CN/HN$^{13}$C; (b)~The kinetic temperature derived using the eq.~\ref{eq:rat_dE};
    (c)~the kinetic temperature derived using CH$_3$CCH transitions and 
    (d)~the kinetic temperature derived using NH$_3$ data. The star-shaped marker and contours are the same as in Fig.~\ref{fig:maps_int_121_30}. 
    The star-shaped marker indicates the IRAS position. The maps of the other sources are presented in the Appendix (Fig.~\ref{fig:maps_Tk}).}
    \label{fig:maps_Tk_121_30}
\end{figure*}

\begin{figure*}
\centering \includegraphics[scale=1.1]{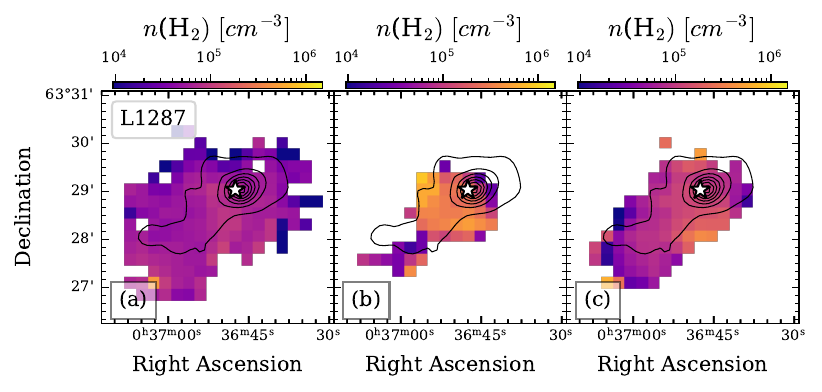}
    \caption{Maps $n($H$_{2})$ for L1287. 
    The H$_{2}$ volume densities are derived using (a)~DCO$^{+}$ line ratio, (b)~DCN line ratio and (c)~DNC line ratio. The star-shaped marker and contours are the same as in Fig.~\ref{fig:maps_int_121_30}. 
    The maps of the other sources are presented in the Appendix (Fig.~\ref{fig:nH2_maps}).} 
    \label{fig:nH2_maps_121_30}
\end{figure*}

The kinetic temperature at the dust emission peak is $\approx 25$~K and decreases to $\approx20$~K towards the southeast along the dust emission ridge~(Figure~\ref{fig:maps_Tk_121_30}). The gas density at the centre is $\sim10^5$\pcmm and slightly decreases to southeast as well as the temperature~(Figure~\ref{fig:nH2_maps_121_30}). 
Based on dust emission the mean density is $3\times 10^4$\pcmm, with a source size of 0.2 pc.

\begin{figure*}
\centering
\centering \includegraphics[scale=1.1]{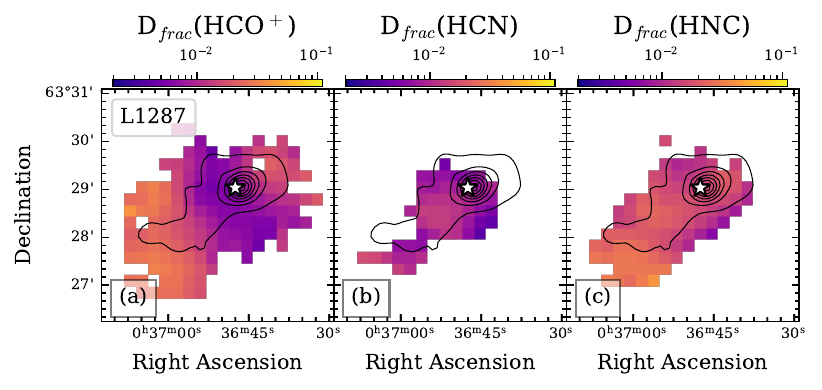}
    \caption{Maps of $D_{\rm frac}$ for L1287. 
    The deuterium fractions are derived using (a)~DCO$^{+}$/HCO$^{+}$ ratios, (b)~DCN/HCN ratios and (c)~DNC/HNC ratios.
    The star-shaped marker and contours are the same as in Fig.~\ref{fig:maps_int_121_30}. The maps of the other sources are presented in the Appendix (Fig.~\ref{fig:Dfrac_maps}).} 
    \label{fig:Dfrac_maps_121_30}
\end{figure*}

\begin{figure}
    \centering \includegraphics[scale=1.1]{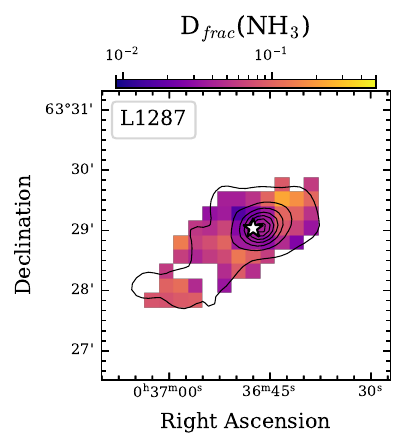}
    \caption{Maps of $D_{\rm frac}(\rm NH_3)$ for L1287.
    The star-shaped marker and contours are the same as in Fig.~\ref{fig:maps_int_121_30}. The maps of the other sources are presented in the Appendix (Fig.~\ref{fig:Dfrac_nh3}).} 
    \label{fig:Dfrac_nh3_121_30}
\end{figure}

The minimum value of the deuterium fraction for $\rm HCO^+$, HCN and HNC is observed towards the center. The $D_{\rm frac}$ are 0.006--0.02 for $\rm HCO^+$, 0.008--0.01 for HCN, 0.015--0.02 for HNC and decrease with rising temperature~(Figure~\ref{fig:Dfrac_maps_121_30}). The $D_{\rm frac}$(NH$_3$) values are 0.03--0.1~(Figure~\ref{fig:Dfrac_nh3_121_30}).

\subsubsection{S187}
The kinetic temperature towards the central dust emission peak is $\approx 20$~K and increases to $\approx25$~K at the IRAS source position~(Figure~\ref{fig:maps_Tk}). The gas density towards the IRAS source position is $\la10^5$\pcmm and is close to the value at the centre position~(Figure~\ref{fig:nH2_maps}).
The mean density is $3\times 10^3$\pcmm, determined towards the central dust emission peak, with a size of 0.3 pc.

Notably, $\rm DCO^+$, unlike DCN and DNC, shows emission at the IRAS source position.
The $D_{\rm frac}$ values at the central peak are 0.04 for $\rm HCO^+$, 0.02 for HCN, and 0.05 for HNC~(Figure~\ref{fig:Dfrac_maps}). The $D_{\rm frac}(\rm HCO^+)$ values at the IRAS source position are 0.02. The mean $D_{\rm frac}$(NH$_3$) value is 0.23 near the central dust emission peak~(Figure~\ref{fig:Dfrac_nh3}).

\subsubsection{S231}
The kinetic temperature at the IRAS source position is 26--30~K. It is close to the value at the central dust emission peak and decreases to $\approx23$~K to the northwest along the dust emission distribution~(Figure~\ref{fig:maps_Tk}). The gas density at the centre is  $\sim10^5$\pcmm, almost throughout the entire region where we see dust emission~(Figure~\ref{fig:nH2_maps}).
The mean density is $10^4$\pcmm, with a source size of 0.4 pc.

There are few D/H values for $\rm HCO^+$ in the centre. The $D_{\rm frac}$ values are 0.003--0.03 for $\rm HCO^+$, 0.004--0.008 for HCN, 0.007--0.035 for HNC and decrease with rising temperature~(Figure~\ref{fig:Dfrac_maps}). The D$_{\rm frac}(\rm NH_3)$ values are 0.03--0.2~(Figure~\ref{fig:Dfrac_nh3}).

\subsubsection{DR21(OH)}
The kinetic temperature at the main dust emission peak is $\approx 30$~K and decreases to $\approx20$~K to the north along the dust emission distribution~(Figure~\ref{fig:maps_Tk}). The gas density is $\la10^6$\pcmm and decreases to $\sim10^5$\pcmm almost throughout dust emission distribution~(Figure~\ref{fig:nH2_maps}).
The mean density is $7\times10^4$\pcmm, with a source size of 0.3 pc.

Notably, the $\rm DCO^+$ line demonstrates weak emission near the main dust emission peak.
The $D_{\rm frac}$ values are 0.004--0.02 for $\rm HCO^+$, 0.008--0.04 for HNC, 0.03--0.25 for NH$_3$ and decrease with rising temperature. The $D_{\rm frac}(\rm HCN)$ are 0.005-0.01 and weakly correlate with temperature~(Figs.~\ref{fig:Dfrac_maps},~\ref{fig:Dfrac_nh3}).

\subsubsection{NGC7538}
The kinetic temperature at the IRAS source position is $\sim 40$~K, at the dust peak in the south it is 30~K and decreases to $\la 25$~K to the southwest~(Figure~\ref{fig:maps_Tk}). The gas density at the dust emission peaks is $\la10^6$\pcmm and decreases to $10^5$\pcmm as well as temperature~(Figure~\ref{fig:nH2_maps}).
The mean densities to the north and south peak are $3\times10^4$\pcmm, with a source size of 0.6 pc.

There are few D/H values for $\rm HCO^+$ observed towards the centre. The $D_{\rm frac}$ values are 0.008 at the dust emission peaks for HNC and increase to 0.02; the D/H values of NH$_3$ are 0.02--0.3 and decrease with rising temperature. The $D_{\rm frac}(\rm HCN)$ values are 0.002--0.005 almost throughout the map. The mean $D_{\rm frac}(\rm HCO^+)$ is 0.005 in the lower temperature region~(Figs.~\ref{fig:Dfrac_maps},~\ref{fig:Dfrac_nh3}).

\subsection{Dependence of deuteration on physical parameters}

\begin{figure*}
\centering
\begin{minipage}{0.33\linewidth}
        \centering \includegraphics[width=\linewidth]{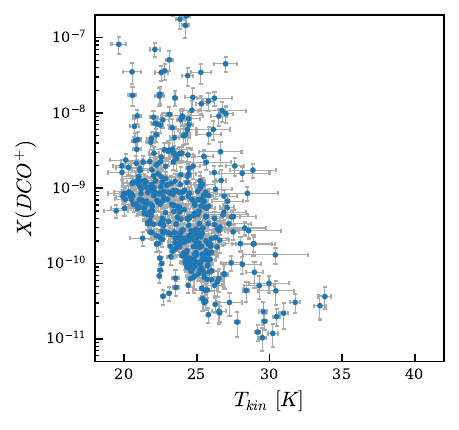}
    \end{minipage}\hfill
    \begin{minipage}{0.33\linewidth}
        \centering \includegraphics[width=\linewidth]{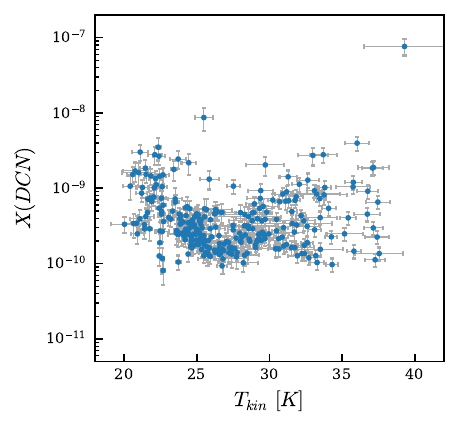}
    \end{minipage}\hfill
    \begin{minipage}{0.33\linewidth}
        \centering \includegraphics[width=\linewidth]{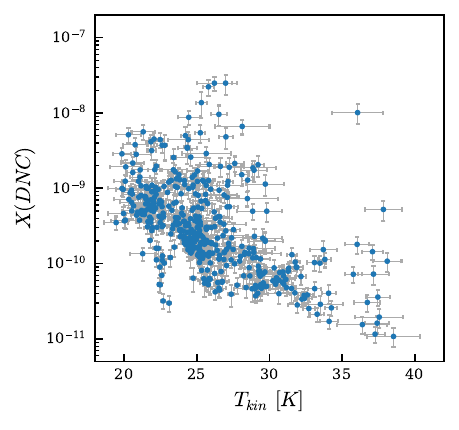}
    \end{minipage}
    \vfill
    \begin{minipage}{0.45\linewidth}
        \centering \includegraphics[width=.75\linewidth]{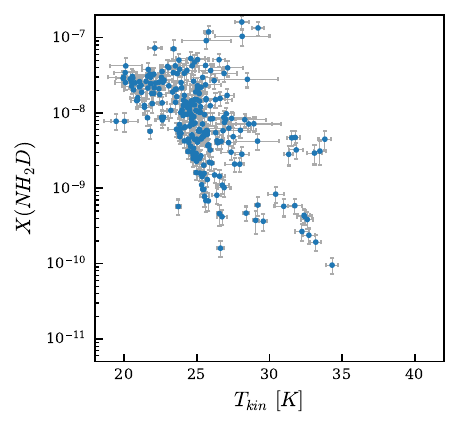}
    \end{minipage}\hfill
    \begin{minipage}{0.45\linewidth}
        \centering \includegraphics[width=.75\linewidth]{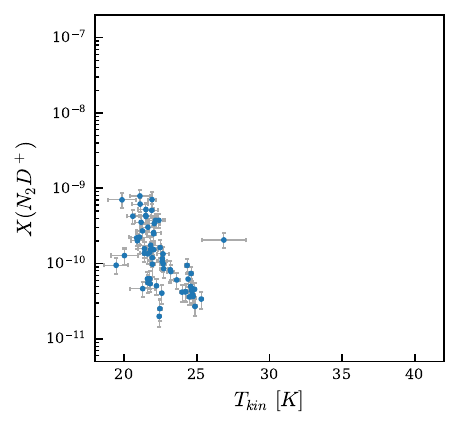}
    \end{minipage}\hfill
    \caption{Dependence of the relative abundances on the kinetic temperature for the DCO$^{+}$, DCN, DNC, NH$_2$D and $\rm N_2D^+$ molecules.} 
    \label{fig:X_mol}
\end{figure*}

We determined abundances of the deuterated molecules relative to H$_2$ (Figure~\ref{fig:X_mol}). These relative abundances are $ \sim 10^{-11}-10^{-9}$ for DCO$^{+}$ and DNC, $ \sim 10^{-11}-10^{-10}$ for $\rm N_2D^+$ and $ \sim 10^{-10}-10^{-8}$ for NH$_{2}$D. The relative abundances of these species decrease with increasing temperature. However, the DCN/H$_2$ ratio is almost constant ($\sim 10^{-10}$).

\begin{figure*}
\centering
    \begin{minipage}{0.33\linewidth}
        \centering \includegraphics[width=\linewidth]{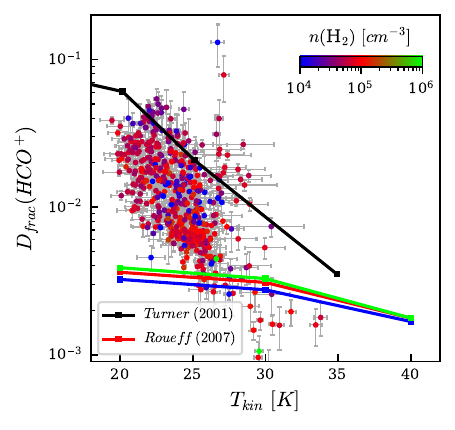}
    \end{minipage}\hfill
    \begin{minipage}{0.33\linewidth}
        \centering \includegraphics[width=\linewidth]{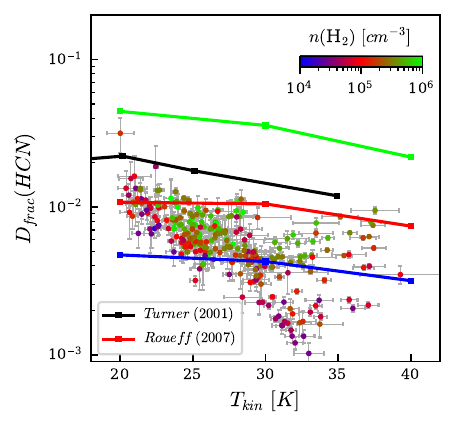}
    \end{minipage}\hfill
    \begin{minipage}{0.33\linewidth}
        \centering \includegraphics[width=\linewidth]{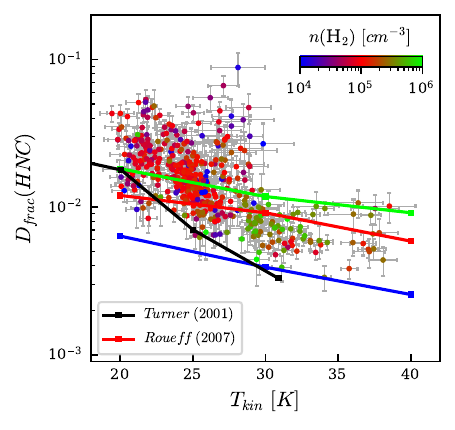}
    \end{minipage}
    \vfill
    \begin{minipage}{0.45\linewidth}
        \centering \includegraphics[width=.75\linewidth]{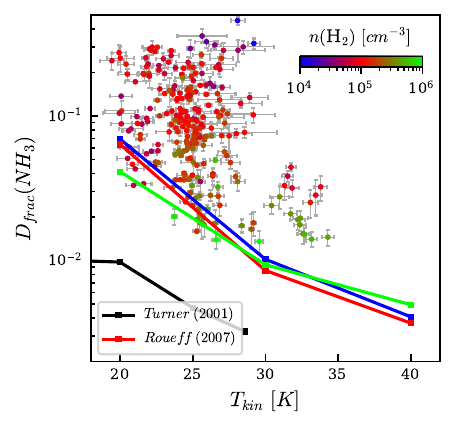}
    \end{minipage}\hfill
    \begin{minipage}{0.45\linewidth}
        \centering \includegraphics[width=.75\linewidth]{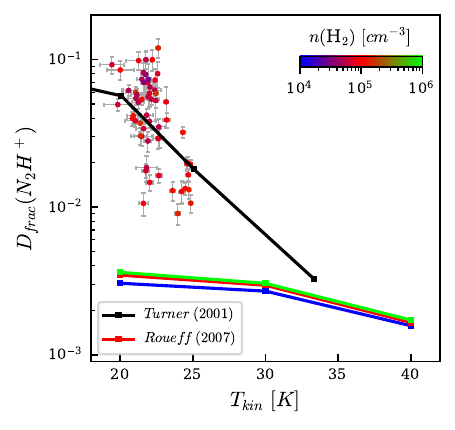}
    \end{minipage}\hfill
    \caption{Dependence of the deuterium fraction on the kinetic temperature for the HCO$^{+}$, HCN, HNC, NH$_3$ and $\rm N_2H^+$ molecules. Color coded bars indicate the gas density values. The blue, red, green and black curves are the model predictions \citep{Roueff07,Turner}.}
    \label{fig:Dfrac}
\end{figure*}

In Fig.~\ref{fig:Dfrac} we show the correlation of the deuterium fraction, the kinetic temperature and the volume density. We also show model predictions by \cite{Roueff07} and \cite{Turner} (see details in section~\ref{literature}). 
We estimate Spearman's rank correlation\footnote{\url{https://docs.scipy.org/doc/scipy/reference/generated/scipy.stats.spearmanr.html}} coefficient $r_{\rm s}$ of D/H ratios, $T_{\rm kin}$ and $n(\rm H_2)$. We define the correlation as "strong" ($|r_{\rm s}|>0.5$), "moderate" ($0.3<|r_{\rm s}|<0.5$) and "weak" ($0.1<|r_{\rm s}|<0.3$).

{$HCO^{+}.$}  The deuterium fraction of HCO$^{+}$ is in the range from 0.001 to 0.05. The $D_{\rm frac}$ and $T_{\rm kin}$ are "strong" anticorrelated at a temperature 20--30~K. The $D_{\rm frac}$ and $n(\rm H_2)$ are "weak" anticorrelated at a volume density of $10^4-10^5$\pcmm. The obtained values are consistent with the model predictions by \cite{Turner}. The model values by \cite{Roueff07} are decreasing more slowly compared to the observational data. 

{\it HCN.} The deuterium fraction of HCN is in the range of 0.001--0.02. The $D_{\rm frac}$ and $T_{\rm kin}$ are "strong" anticorrelated. The correlation of $D_{\rm frac}$ and $n(\rm H_2)$ is "moderate". The $n(\rm H_2)$ is in the range of $10^4-10^6$\pcmm. The obtained values agree with the models data with densities $\la 10^5$\pcmm.

{\it HNC.} The deuterium fraction of HNC is in the range of 0.002--0.05. The $D_{\rm frac}-T_{\rm kin}$ and $D_{\rm frac}-n(\rm H_2)$ values are "strong" anticorrelated at a temperature of 20--40~K and $n(\rm H_2) \sim 10^4-10^6$\pcmm. The obtained values agree with the model curves.

{$NH_3.$} The deuterium fraction of NH$_3$ decreases from 0.4 to 0.02. The $D_{\rm frac}$-$T_{\rm kin}$ values are "moderate" anticorrelated, while $D_{\rm frac}$-$n(\rm H_2)$ are "strong" anticorrelated. The obtained values are higher than the model predictions by \cite{Turner}. The model values by \cite{Roueff07} agree with the observational data.

{$N_2H^{+}.$} There are few D/H values for $\rm N_2H^{+}$. The deuterium fraction of $\rm N_2H^{+}$ is 0.008--0.1 at temperatures 20--25~K, at a density $\sim 10^5$\pcmm. The observed values show similar trends to the model by \cite{Turner}, but are larger than those of the model by~\cite{Roueff07}. 

\section{Discussions}\label{dis}

\subsection{Comparison of the different estimates of the H$_2$ volume density}

We obtained the H$_2$ volume densities using both the dust continuum emission and non-LTE analysis. The densities derived from the molecular excitation analysis ($\sim 10^5$\pcmm) are on average an order of magnitude higher than the mean densities ($\sim 10^4$\pcmm). This is a typical situation \citep[e.g.,][]{Zinchenko1994,Zinchenko1998}, which can be probably explained by the small-scale clumpiness in the sources.

\subsection{The ortho-to-para ratio of NH$_2$D}

To derive the total column density of NH$_2$D we were using o-NH$_2$D lines assuming the o-/p- ratio of 3:1. The p-NH$_2$D line at 110~GHz was also detected in DR21(OH). We decided to verify whether the assumption is correct. For this, integrated intensity o-/p- ratios were obtained, and as a result, the mean value is 2.6, with a standard deviation of 0.7. 
\cite{Fontani15} found similar results for high-mass star-forming samples covering different evolutionary phases. They compared integrated intensities and derived a mean value of $2.6 \pm 0.6$.  
\cite{Wienen2021} determined a median ortho-to-para column density ratio of $3.7 \pm 1.2$ from the ATLASGAL survey. They also derived this value from integrated intensity ratio. This resulted in a mean value of $2.6 \pm 0.8$, which is close to our results. 
Additionally, \cite{Sipila19} constructed models for deuterium and spin-state chemistry for the purpose to model the low-temperature environment in starless and pre-stellar cores. They found, the models presented an NH$_2$D ortho-to-para ratio of 2.

\subsection{Comparison of deuteration with other results}\label{literature}

Here we review some of the modelling and observational main work, and highlight the differences with our results.
To study the chemistry of deuterated species \cite{Turner} used a model, containing 9930 reactions and 610 species. The model described the dependence of the molecular D/H ratios upon temperature, density, ionization rate, extinction, epoch, and elemental abundances. At temperature 20--35~K and density 10$^5$\pcmm the predicted molecular D/H ratios are shown in Fig.~\ref{fig:Dfrac}.
\cite{Roueff05} presented a steady state model of the gas-phase chemistry. At temperature 20--50~K and density 10$^5$\pcmm the molecular D/H ratios decrease from 0.1 to 0.002 for the discussed species. 
After that in \cite{Roueff07} the abundance ratios were analysed as functions of temperature and density for the “standard” low-metal set, which leads to best results for dark cold clouds, and the “warm-core” set, in which higher values of heavy elements are used to account for partial evaporation of grain mantles. These “warm-core” models are illustrated in Fig.~\ref{fig:Dfrac}.

\cite{Albertsson} calculated the deuterium fractionation model and simulated it in diverse interstellar sources. For prestellar objects the model predicted 0.01--1 for DCO$^+$/HCO$^+$ and  $\rm N_2D^+$/$\rm N_2H^+$, as well as 0.001--0.01 for DCN/HCN, DNC/HNC and NH$_2$D/NH$_3$. The DCO$^+$ and $\rm N_2D^+$ values are greater than our results. Also the model prediction of NH$_2$D/NH$_3$ is less than our observational data.  
\cite{Sipila15} developed the gas-grain models and the deuterium fraction of ammonia was $\sim$0.1 after 10$^5$ yr at a density of 10$^5$\pcmm and a temperature of 15 K.
\cite{Sipila19} presented a chemical model for deuterium chemistry. They found that the highest ammonia D/H ratios were 0.1--1 at $\sim 10^5$\pcmm and after 10$^5$ yr.

\cite{Sakai12} conducted a survey towards 18 massive clumps. For the five high-mass protostellar objects at temperature $\sim$25~K the mean DNC/HNC ratio was 0.0095. 
\cite{Fontani11} found that $\rm N_2D^+$/$\rm N_2H^+$ and $T_{\rm kin}$ are slightly anti-correlated. The mean $D_{\rm frac}$ decreases from $\sim$0.26 in the HMSCs to $\sim$0.04 in the HMPOs and UCHIIs. Their results in the HMSCs with temperature below 20~K are greater than our observed data.  
\cite{Fontani14} observed DNC(1--0) and HN$^{13}$C(1--0) towards 22 massive star-forming cores. They found an average $D_{\rm frac}$(HNC) of 0.01 with no significant correlation between the three evolutionary groups of sources.
\cite{Fontani15} observed NH$_2$D and NH$_3$ towards previously observed massive star-forming regions.  $D_{\rm frac}$(NH$_3$) was 0.01--1 at temperature 10--30~K and does not change significantly from the earliest to the most evolved phases. Additionally, for S231, values of 0.498 and 0.191 were found.

\cite{Gerner15} observed a sample of 59 sources including different evolutionary groups. They found  D/H ratios of 0.0004--0.02 for DCO$^+$, 0.003--0.03 for DCN, 0.001--0.02 for DNC and 0.001--0.01 for $\rm N_2D^+$. They also reported that the D/H ratios of DNC, DCO$^+$ and $\rm N_2D^+$ decrease with time. DCN/HCN peaks at the HMC stage. {They also modelled and observed relative abundances to H$_2$. The observed values were $\sim 10^{-11}$ of $X$(DCO$^{+}$), $X$(DCN), $X$(DNC) and $X$($\rm N_2D^+$), mainly close to the model results. However, the model values of $X$($\rm N_2D^+$) were $10^3$ less then indicated by observations of the infrared dark cloud phase (IRDC). Our} obtained relative abundances are greater by an order of magnitude.

\cite{Feng19} obtained maps towards young high-mass star-forming clumps in G28.34+0.06. As the temperature decreases from 20 to 14~K, the molecular D/H ratios become 0.03--0.06 for $\rm N_2D^+$, 0.004--0.006 for DCO$^+$, 0.006--0.01 for DCN, 0.008--0.013 for DNC and 0.01--0.005 for NH$_2$D. Our estimates at $\sim$20~K are greater then these results.
\cite{Trofimova20} have carried out a survey of 60 massive star forming regions (including our sources) of DCN, DNC, DCO$^+$, $\rm N_2D^+$, by using the 20-m Onsala radio telescope. The CH$_3$CCH $J=5-4$ transitions were used to estimate the rotational temperature. For L1287, they obtained 0.065 for DCO$^+$ and 0.017 for DCN at 34.4~K, 0.007 for DCN at 40.2~K in S231, 0.01 for DCN at 29.6~K in DR21(OH), 0.007 for DCN at 47.7~K in NGC7538. These values are comparable to our results. However, our estimates of temperature are 10~K less than theirs in L1287 and S231. 

\begin{figure}
    \centering \includegraphics[width=\linewidth]{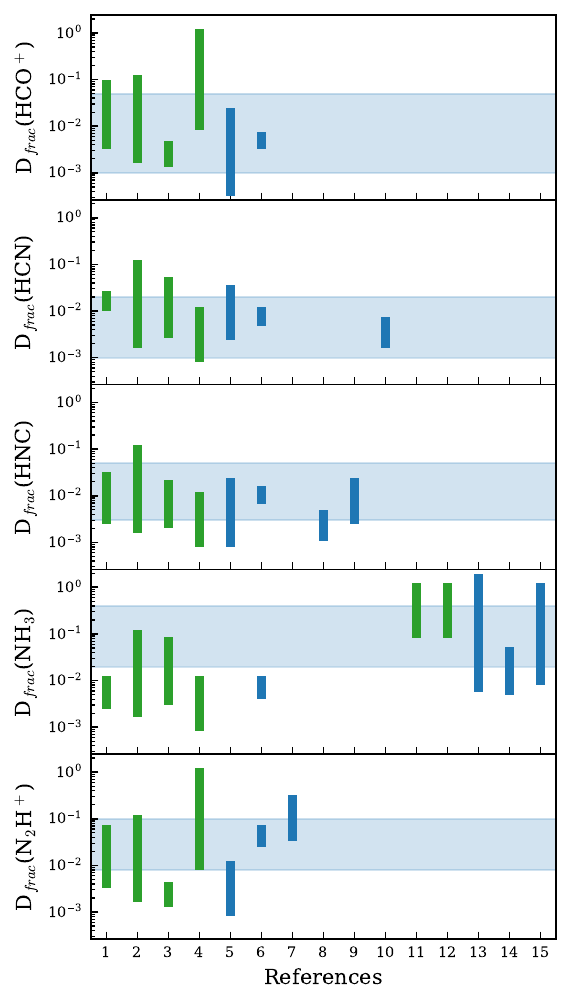} 
    \caption{Comparison of observed D/H values with previous works. The blue and green bars show the observational and chemical model results, respectively. The blue shaded region indicates values from this work.
    References: (1)~\citet{Turner}; (2)~\citet{Roueff05}; (3)~\citet{Roueff07}; (4)~\citet{Albertsson}; (5)~\citet{Gerner15}; (6)~\citet{Feng19}; (7)~\citet{Fontani11}; (8)~\citet{Sakai12}; (9)~\citet{Fontani14}; (10)~\citet{Trofimova20}; (11)~\citet{Sipila15}; (12)~\citet{Sipila19}; (13)~\citet{Wienen2021}; (14)~\citet{Li22}; (15)~\citet{Fontani15}. }
    \label{summary}
\end{figure}

\cite{Wienen2021} observed a sample of high-mass clumps discovered by the ATLASGAL survey covering various evolutionary phases of massive star formation. The NH$_2$D-to-NH$_3$ column density ratios range from 0.007 to 1.6. They did not find a correlation between the NH$_3$ deuteration and evolutionary tracer such as rotational temperature in the range 10--25~K.
\cite{Li22} observed NH$_2$D at 110 GHz towards 50 Galactic massive star-forming regions.
An excitation temperature $T_{\rm ex}$ of 18.75 K was used to calculate column densities. The range of deuterium fractionation encompasses values from 0.043 to 0.0006. The D/H ratio is 0.043 in DR21 and 0.015 in NGC7538. These results are close to our results.
We summarise the observed D/H values and the results of previous works in Fig. \ref{summary}.

\section{Conclusions}\label{con}

Using observations with the IRAM-30m radio telescope and the 100-m radio telescope in Effelsberg, we have obtained the spatial distributions of the $J=1-0$ and $J=2-1$ DCO$^{+}$, DCN, DNC, $\rm N_2D^+$ lines and $1_{11}-1_{01}$ ortho- and para-NH$_{2}$D lines in five high-mass star-forming regions. We have derived deuterium fractions as functions of gas properties such as temperature and density. This has been combined with H$_2$ column densities from $850\,\micron$ SCUBA dust continuum maps. 

The results are as follows:
\begin{enumerate}

\item[1.]  The deuterated molecules suggest different spatial distributions. DCN, unlike DNC, DCO$^+$ and NH$_2$D, shows emission peaks consistent with the hydrogenated isotopologues. Notably, in S187, the NH$_2$D line demonstrates stronger emission by a factor of $\sim 3$ than NH$_3$, but it is located at the edge of the map. The $\rm N_2D^+$ emission is weak in most of the sources.
 
\item[2.] 
The kinetic temperature was estimated from CH$_3$CCH and NH$_3$ lines, as well as using the integrated intensity ratios of the $J=1-0$ H$^{13}$CN and HN$^{13}$C lines and their main isotopologues. The T$_{\rm kin}$(HCN/HNC) maps show a good agreement with the estimates derived from the CH$_3$CCH and NH$_3$ lines in the range of 20 to 40 K.
Using the $J$=2--1/1--0 integrated line ratios and the T$_{\rm kin}$(HCN/HNC) with the RADEX code, we have estimated the gas density. Densities are $\sim 10^4-10^6$\pcmm. 

\item[3.] The abundances relative to H$_2$ are $\sim 10^{-9}-10^{-11}$ for DCO$^{+}$ and DNC, $\sim 10^{-10}-10^{-11}$ for $\rm N_2D^+$ and $\sim 10^{-8}-10^{-10}$ for NH$_{2}$D. The relative abundances of these species decrease with increasing temperature. However, the DCN/H$_2$ ratio is almost constant ($\sim 10^{-10}$).

\item[4.] We calculated the total column density of deuterated molecules and their hydrogenated isotopologues to determine the deuterium fraction. The $D_{\rm frac}$ are 0.001--0.05 for DCO$^{+}$, 0.001--0.02 for DCN, 0.001--0.05 for DNC and 0.02--0.4 for NH$_{2}$D. The D/H ratios decrease with increasing temperature in the range of 20--40~K and slightly vary at $n(\rm H_2) \sim 10^4-10^6$\pcmm. The deuterium fraction of $\rm N_2H^{+}$ is 0.008--0.1 in the temperature range of 20--25~K and at the density $\sim 10^5$\pcmm.

In addition, we compared those results with the model predictions and observations from the literature (see Figs.~\ref{fig:Dfrac},~\ref{summary}). The observational results agree with the predictions of chemical models (although in some cases there are significant differences). The $D_{\rm frac}$ values range is mostly consistent with those found in other works.

\item[5.] In DR21(OH), para-NH$_2$D at 110~GHz and ortho-NH$_2$D at 86~GHz were detected. We derived the integrated intensity ortho-to-para ratio of NH$_2$D. As a result, the mean value is 2.6, with standard deviation 0.7. 

\end{enumerate}

\section*{Acknowledgements}
This study was supported by the Russian Science Foundation grant No. 22-22-00809.
The research is based on observations made by the 041-19 project with the 30-m telescope, as well as observations with the 100-m MPIfR telescope (Max-Planck-Institut für Radioastronomie) in Effelsberg. IRAM is supported by INSU/CNRS (France), MPG (Germany) and IGN (Spain). We acknowledge the staff of both observatories for their support in the observations. The authors are grateful to the anonymous reviewer for useful comments that improved the quality of the manuscript.

\section*{Data Availability}

The original data obtained with IRAM-30m are available under the IRAM Data Archive. The data underlying this article will be shared on reasonable request to the corresponding author.



\bibliographystyle{mnras}
\bibliography{pazukhin} 




\appendix

\section{Figures}
This section contains additional figures for S187, S231, DR21(OH) and NGC7538.
Figure~\ref{fig:maps_int} illustrates the integrated intensity maps. 
Figure~\ref{fig:maps_Tk} shows the kinetic temperature maps.
Figure~\ref{fig:nH2_maps} represents the volume gas density $n($H$_{2})$ maps.
Figures~\ref{fig:Dfrac_maps}~and~\ref{fig:Dfrac_nh3} visualise maps of the deuterium fraction $D_{\rm frac}$ for HCO$^+$, HCN, HNC and NH$_3$.

\begin{figure*}
    \centering \includegraphics[width=\linewidth]{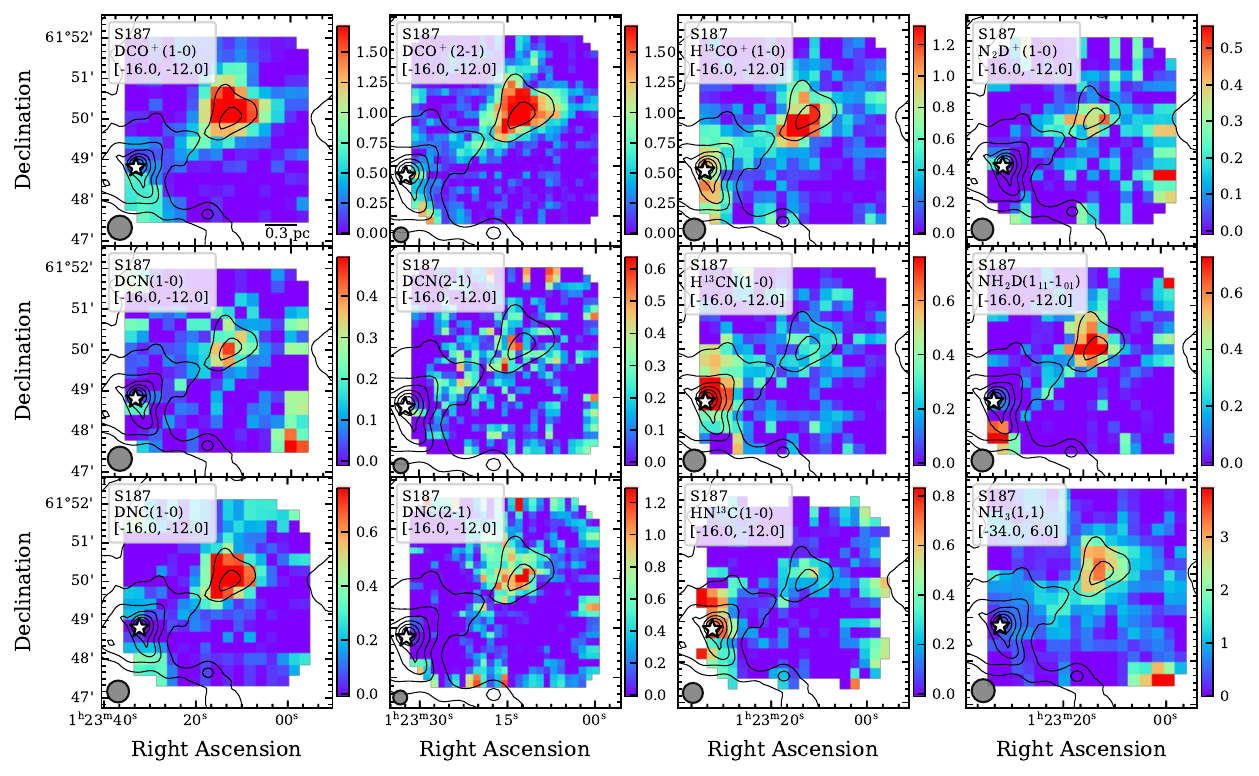}  
    \caption{The integrated intensity maps. The integrated intensity is shown in units of main-beam temperature [K\kms]. The contours show the continuum emission from 850\, $\micron$ SCUBA data. The levels start from 5\% to 95\% of the peak intensity of 2.9 $\rm mJy\, beam^{-1}$ in steps of 15\%. The star-shaped marker indicates the IRAS source position. Sources, transitions  and the velocity range are shown in the upper left corner of each panel. The beam sizes are shown in the bottom left corner of each panel. A scale bar representing a linear scale of 0.3 pc is shown on the bottom-right corner of the first frame.}
    \label{fig:maps_int}
\end{figure*}

\addtocounter{figure}{-1}
\begin{figure*}
    \centering \includegraphics[width=\linewidth]{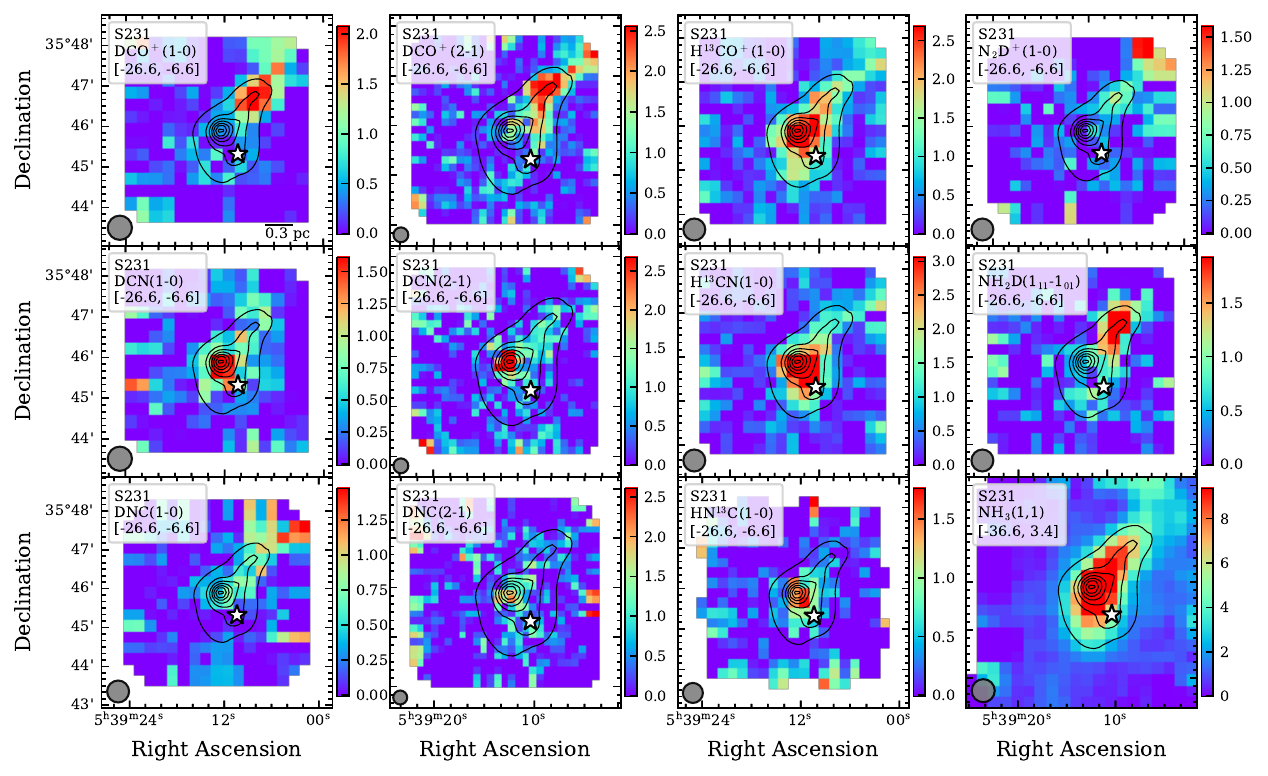}
    \caption{continued. The peak intensity is 8.3 $\rm mJy\, beam^{-1}$.}
\end{figure*}
\addtocounter{figure}{-1}
\begin{figure*}
    \centering \includegraphics[width=\linewidth]{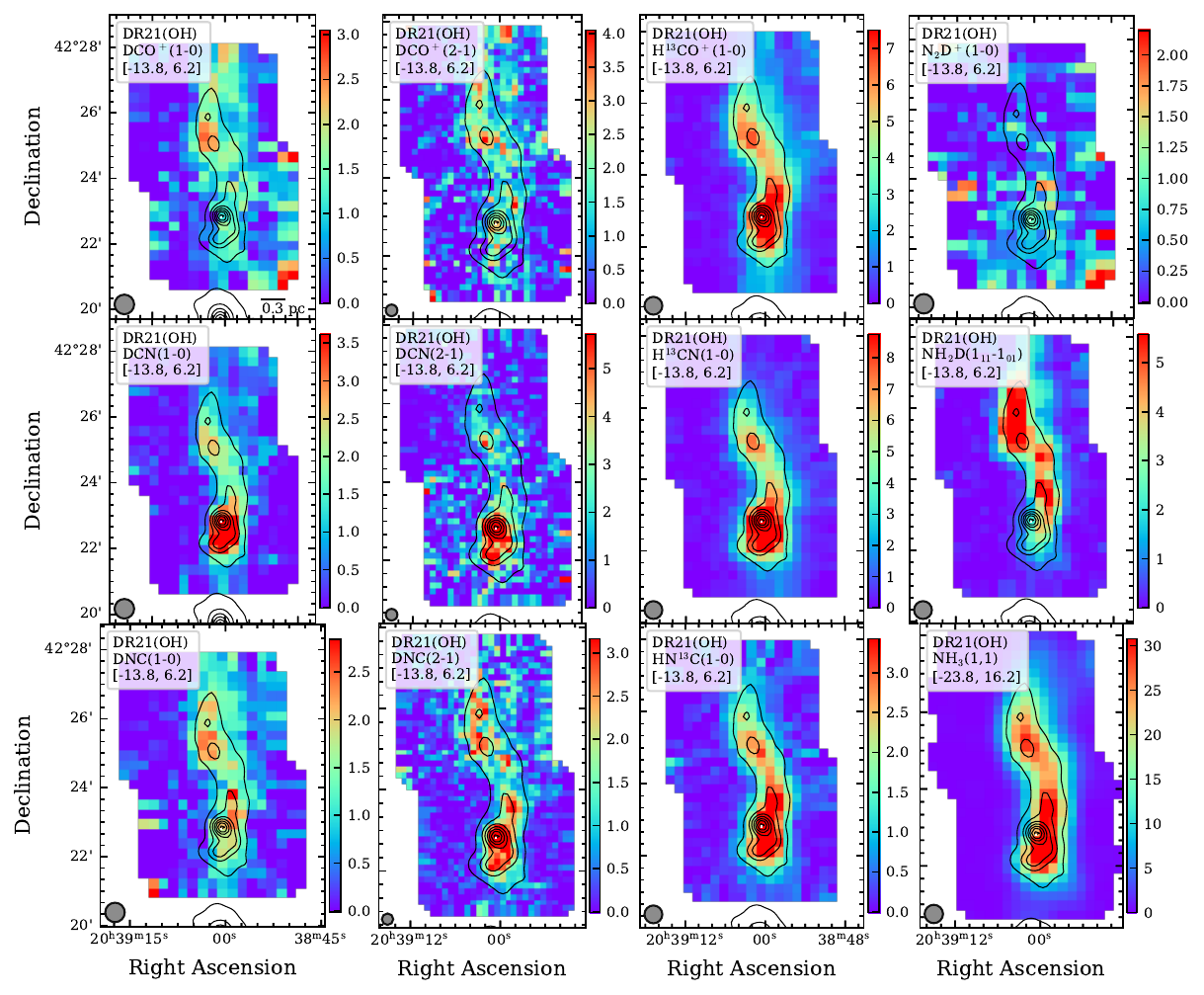}
    \caption{continued. The peak intensity is 33.7 $\rm mJy\, beam^{-1}$.}
\end{figure*}
\addtocounter{figure}{-1}
\begin{figure*}
    \centering \includegraphics[width=\linewidth]{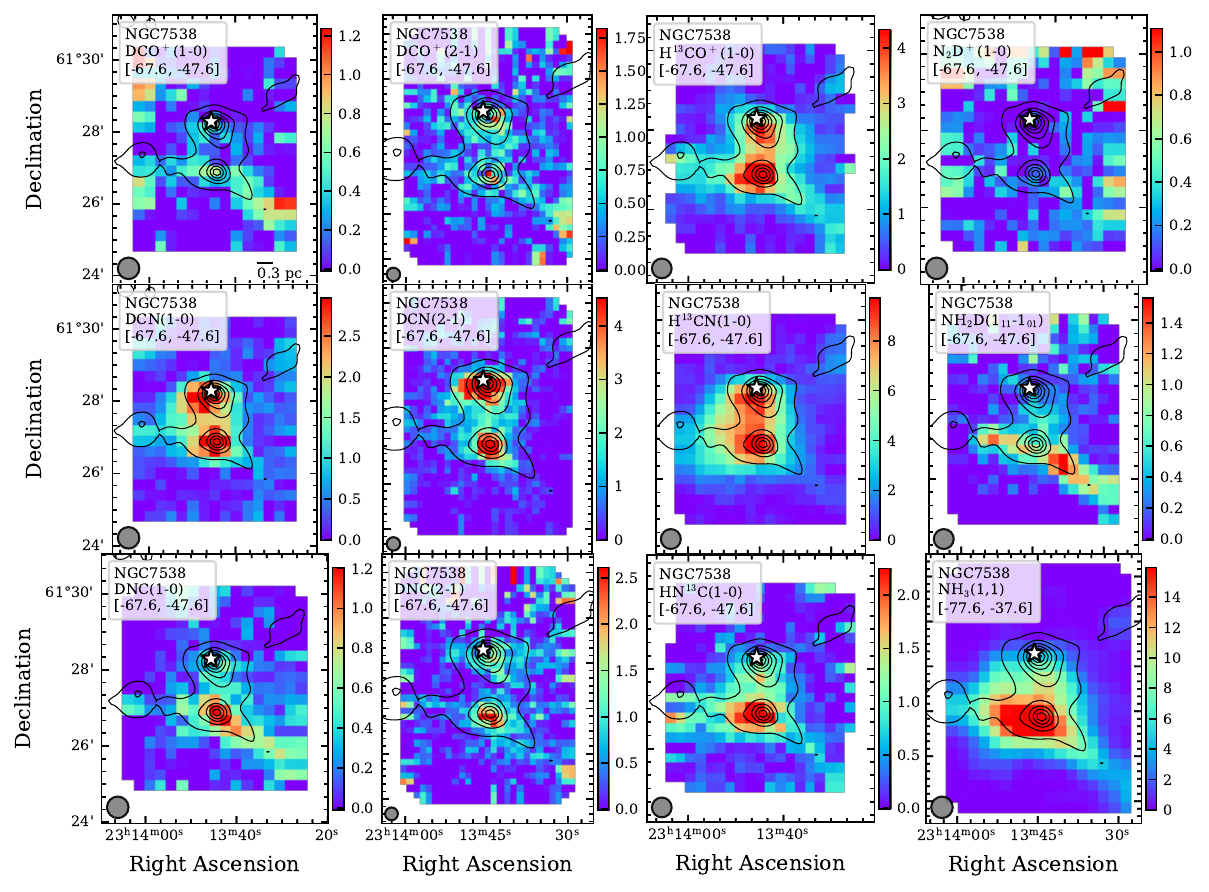}    
    \caption{continued. The peak intensity is 23.4 $\rm mJy\, beam^{-1}$.}
\end{figure*}

\begin{figure*}
    \centering \includegraphics[width=\linewidth]{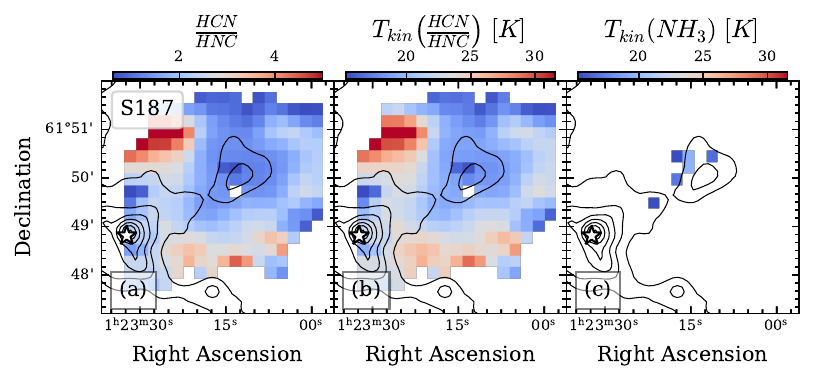}
    \centering \includegraphics[width=\linewidth]{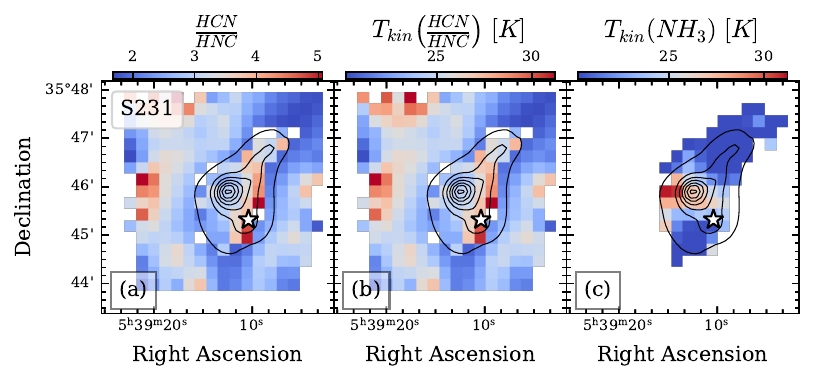}
    \caption{Maps for S187 (upper panels) and S231 (lower panels).  (a)~Combined integrated intensity ratios HCN/HNC and H$^{13}$CN/HN$^{13}$C; (b)~the kinetic temperature derived using the eq.~\ref{eq:rat_dE};
    (c)~the kinetic temperature derived using NH$_3$ transitions. 
    The star-shaped marker and contours are the same as in Fig.~\ref{fig:maps_int}.} 
    \label{fig:maps_Tk}
\end{figure*}

\addtocounter{figure}{-1}
\begin{figure*}
    \centering \includegraphics[width=\linewidth]{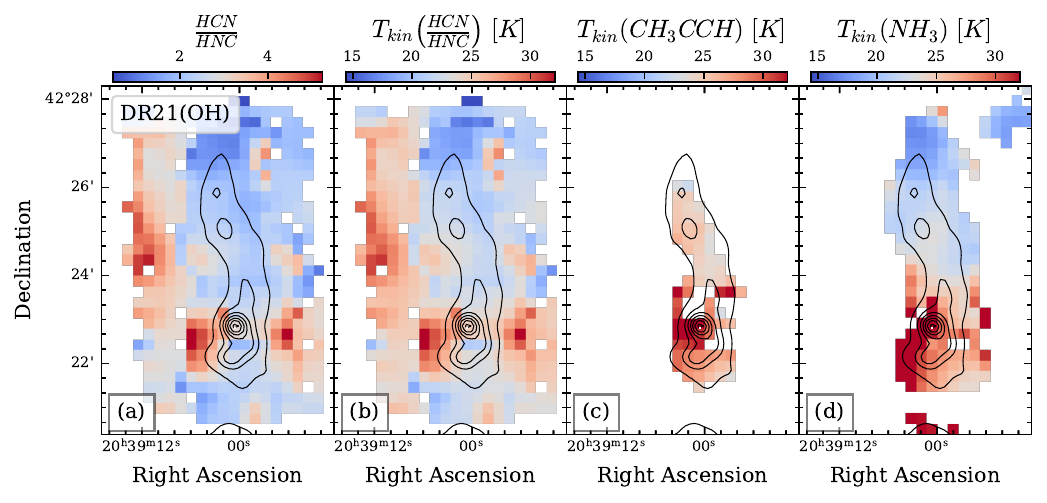}
    \centering \includegraphics[width=\linewidth]{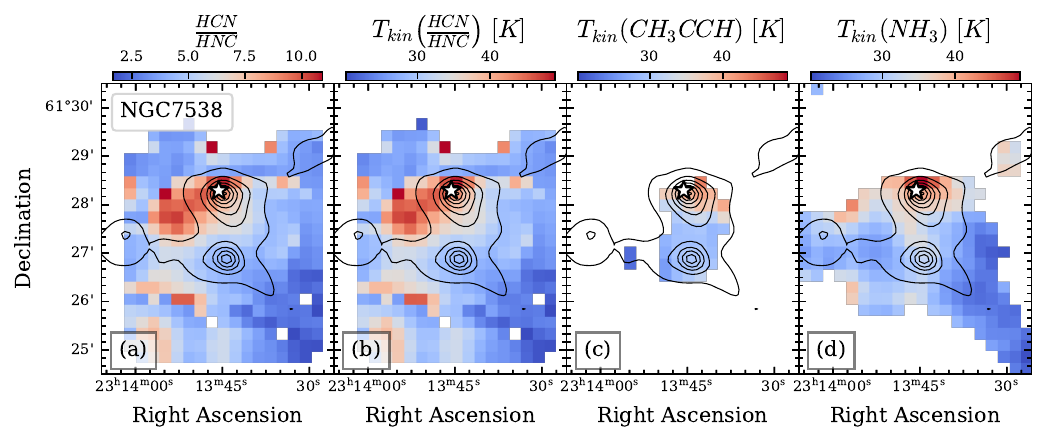}
    \caption{continued. (c)~the kinetic temperature derived using CH$_3$CCH transitions. (d)~The kinetic temperature derived using NH$_3$ data.}
\end{figure*}

\begin{figure*}
\centering
\centering \includegraphics[width=\linewidth]{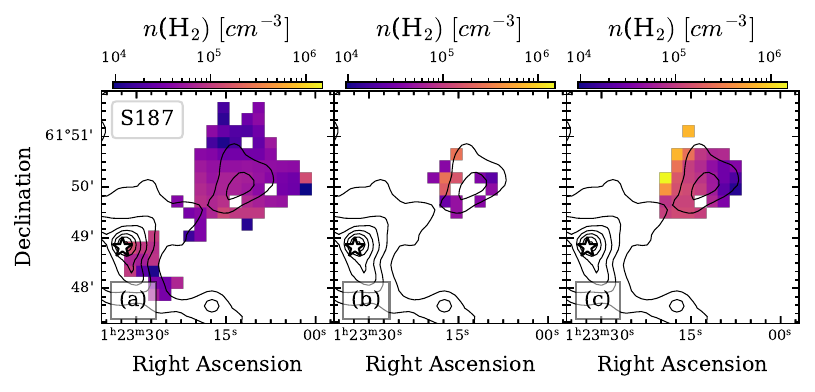}
\centering \includegraphics[width=\linewidth]{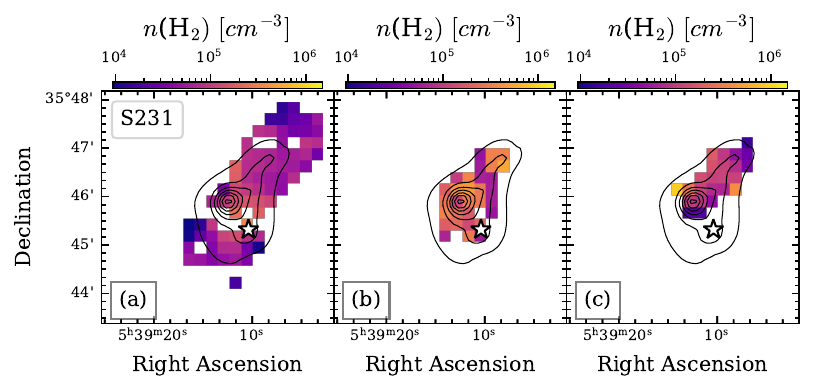}
    \caption{Maps $n($H$_{2})$ for S187 (upper panels) and S231 (lower panels).  
    The H$_{2}$ volume densities are derived using (a)~DCO$^{+}$ line ratios, (b)~DCN line ratios and (c)~DNC line ratios.
    The star-shaped marker and contours are the same as in Fig.~\ref{fig:maps_int}. } 
    \label{fig:nH2_maps}
\end{figure*}

\addtocounter{figure}{-1}
\begin{figure*}
\centering
\centering \includegraphics[width=\linewidth]{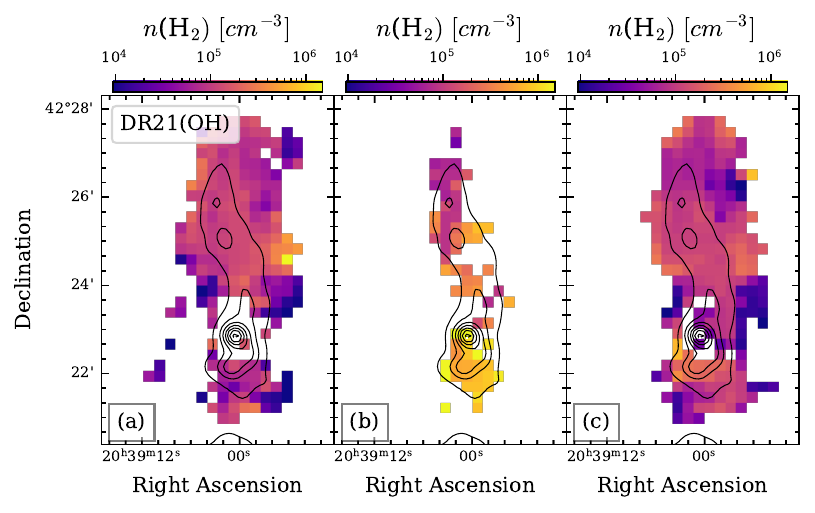}
\centering \includegraphics[width=\linewidth]{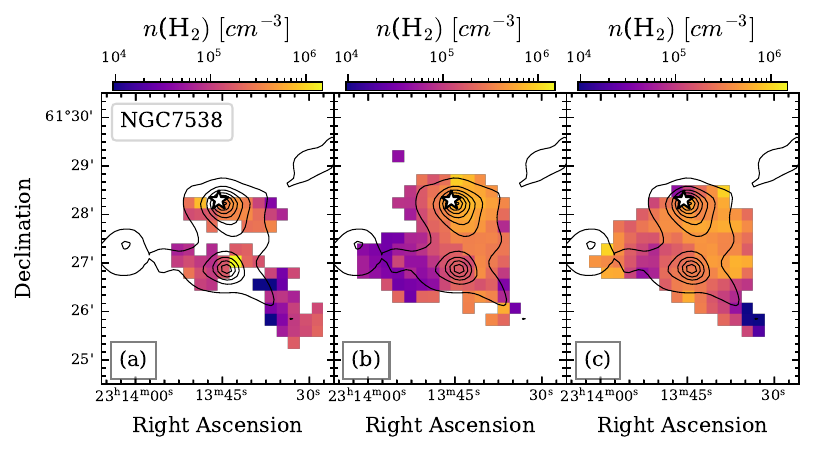}
    \caption{continued.}
\end{figure*}

\begin{figure*}
\centering
\centering \includegraphics[width=\linewidth]{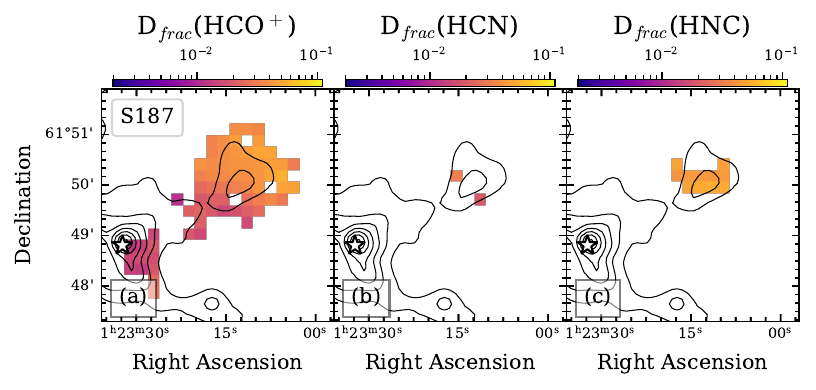}
\centering \includegraphics[width=\linewidth]{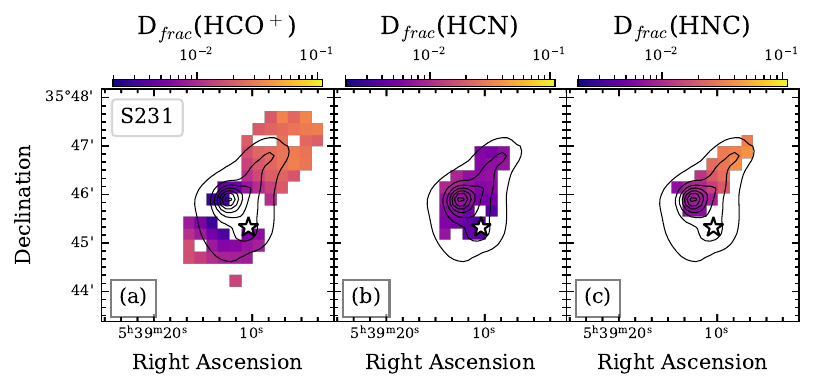}
    \caption{Maps of $D_{\rm frac}$ for S187 (upper panels) and S231 (lower panels).
    The deuterium fractions are derived using (a)~DCO$^{+}$/HCO$^{+}$ ratios, (b)~DCN/HCN ratios and (c)~DNC/HNC ratios.
    The star-shaped marker and contours are the same as in Fig.~\ref{fig:maps_int}. } 
    \label{fig:Dfrac_maps}
\end{figure*}

\addtocounter{figure}{-1}
\begin{figure*}
\centering
\centering \includegraphics[width=\linewidth]{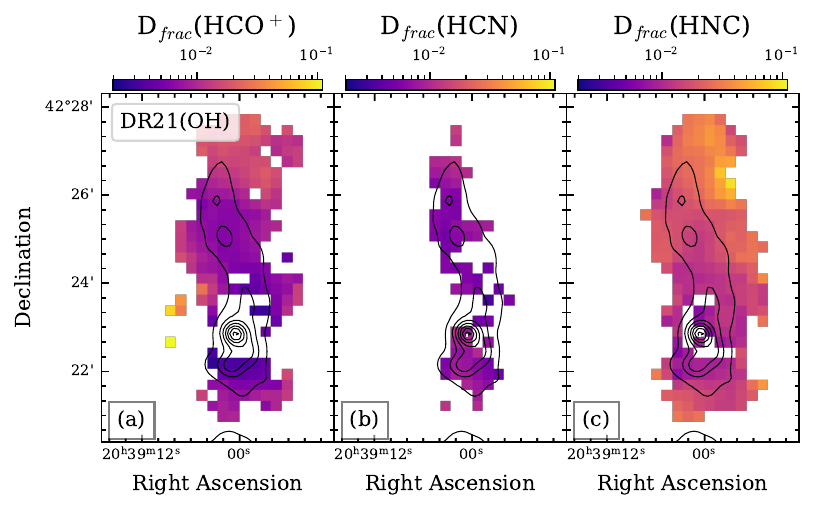}
\centering \includegraphics[width=\linewidth]{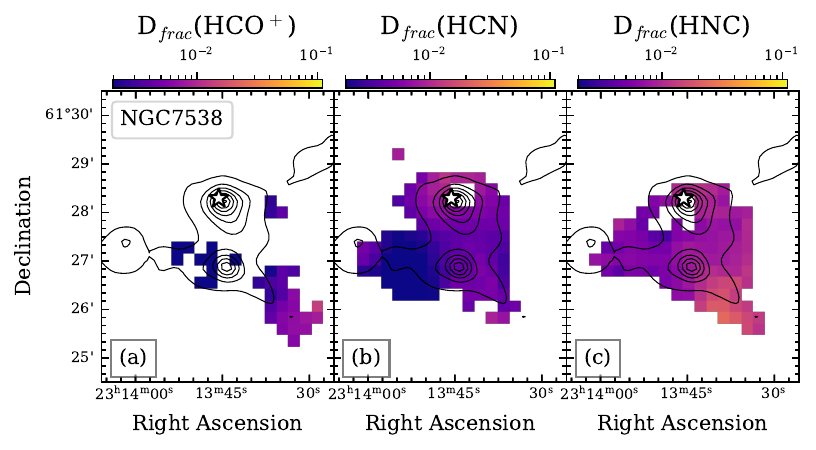}
    \caption{continued.}
\end{figure*}

\begin{figure*}
\centering
    \begin{minipage}{0.45\linewidth}
        \centering \includegraphics[width=\linewidth]{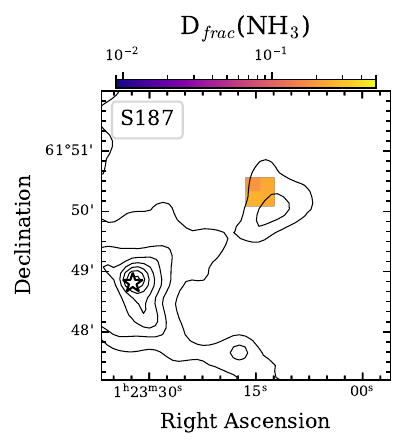}
    \end{minipage}
    \begin{minipage}{0.45\linewidth}
        \centering \includegraphics[width=\linewidth]{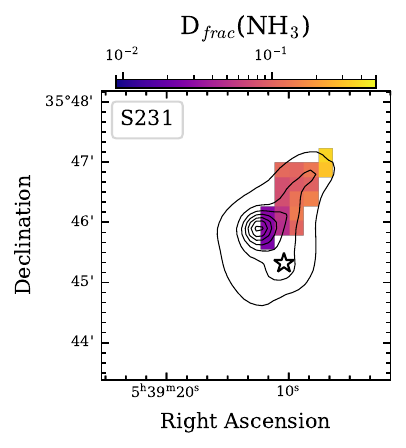}
    \end{minipage}
    \vfill
        \begin{minipage}{0.45\linewidth}
        \centering \includegraphics[width=\linewidth]{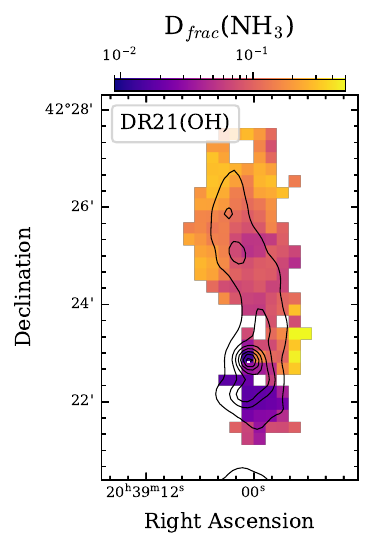}
    \end{minipage}
    \begin{minipage}{0.45\linewidth}
        \centering \includegraphics[width=\linewidth]{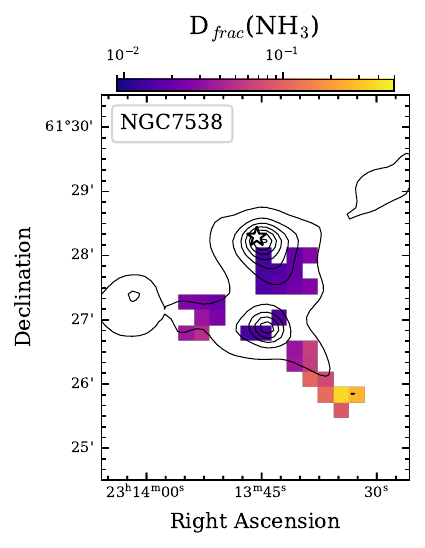}
    \end{minipage}    
    \caption{Maps of $D_{\rm frac}(\rm NH_3)$ for S187, S231, DR21(OH) and NGC7538.
    The star-shaped marker and contours are the same as in Fig.~\ref{fig:maps_int}.} 
    \label{fig:Dfrac_nh3}
\end{figure*}


\bsp	
\label{lastpage}
\end{document}